\documentclass{aa}

\usepackage{graphicx}
\usepackage{txfonts}
\usepackage{natbib}
\usepackage{newtxtext,newtxmath}
\usepackage{amsmath}
\usepackage{color}
\usepackage{multirow}

\newcommand{\Mdot}{\rm {M}_{\odot}}
\newcommand{\kms}{{\rm {km~s}}^{-1}}
\usepackage[normalem]{ulem}

\begin{document}

\title{Velocity dispersion functions of pressure-supported galaxies in EAGLE simulations with varying active galactic nucleus feedback}
\titlerunning{VDFs of Pressure-supported Galaxies in EAGLE Simulations with Varying AGN Feedback}

\author{Jungwon Choi
          \inst{1}
          \and
        Jubee Sohn\inst{1, 2}}
\authorrunning{J. Choi and J. Sohn}

\institute{Department of Physics and Astronomy, Seoul National University, 1 Gwanak-ro, Gwanak-gu, Seoul 08826,
Republic of Korea\\
\email{jwchoi29912@gmail.com}
\and
SNU Astronomy Research Center, Seoul National University, Seoul 08826, Republic of Korea\\
\email{jubee.sohn@snu.ac.kr}}
 
\abstract
{We investigated the stellar velocity dispersion functions (VDFs) of pressure-supported galaxies in the EAGLE cosmological simulations. The central stellar velocity dispersion is one of the fundamental dynamical tracers of the total mass of galaxy subhalos, alongside luminosity and stellar mass. Because it reflects the gravitational potential, the stellar velocity dispersion is expected to be relatively insensitive to feedback from active galactic nuclei (AGNs), a critical process that regulates the connection between other galaxy observables and subhalo masses. To examine the impact of AGN feedback, we analyzed the VDFs from five EAGLE simulation runs, each adopting a different AGN feedback model: one "standard", two "enhanced", one "reduced," and one with no AGN feedback. We computed the stellar velocity dispersions of pressure-supported galaxies using member stellar particles, mimicking fiber spectroscopy. The VDFs from the standard and enhanced AGN feedback models show little difference. However, contrary to our initial expectation that the VDF shape would be largely insensitive to AGN feedback, the simulations with reduced and no AGN feedback show a significant excess of high velocity dispersion galaxies ($\sigma_{*} > 200~\kms$) and a deficit of low velocity dispersion galaxies ($100 < \sigma_{*} (\kms) < 200$), compared to those with standard or enhanced AGN feedback. The presence of high velocity dispersion galaxies in the no-AGN model arises from enhanced central star formation, due to the absence of AGN-driven gas heating or expulsion. Our results demonstrate that the shape of the theoretical VDF is sensitive to the strength of AGN feedback. These predictions offer a theoretical benchmark for future observational studies of the galaxy VDF using large-scale spectroscopic surveys.}


\maketitle

\section{Introduction} \label{sec:intro}

A standard $\Lambda$ Cold Dark Matter (CDM) model suggests that the mass concentration of dark matter (DM) determines the structure formation. Within the DM halos, the gravitational interaction between DM and baryonic matter initiates the complicated formation and evolution of galaxies therein. One of the greatest challenges in studying galaxy evolution in DM halos is identifying an observable that connects the underlying DM halos and observable baryonic matter.

The central stellar velocity dispersion of a galaxy is one of the fundamental galaxy observables that connect the galaxy properties and the underlying DM mass. The central velocity dispersion is a kinematic measure determined by the central gravitational potential of galaxies defined by the total mass of galaxy (sub)halos including DM. Stellar velocity dispersions show good correlations with other observables including luminosity or stellar mass (e.g., \citealp{Faber1976, Djorgovski1987, Cappellari2013a, Mogotsi2019, Cannarozzo2020, Ferragamo2021}). More importantly, various scaling relations based on the stellar velocity dispersion indicate that stellar velocity dispersions are an excellent tracer of the total mass of galaxy subhalos. For example, the stellar velocity dispersions show a tight correlation with the DM velocity dispersion or the DM mass of subhalos \citep{Wake2012, vanUitert2013, Bogdan2015, Zahid2016, Zahid2018, Sohn2024a}. The observational studies based on strong or weak gravitational lensing also demonstrate that the stellar velocity dispersion traces the total mass of galaxy subhalos (e.g., \citealp{Grillo2008, Auger2010, Utsumi2020}).

The stellar velocity dispersion has multiple advantages in tracing the total mass of galaxies. First of all, stellar velocity dispersion is relatively straightforward to obtain compared to other mass tracers including galaxy luminosity or stellar mass. These conventional mass tracers, galaxy luminosity or stellar mass, based on photometry are often suffering from the contamination by the line-of-sight objects or uncertain boundaries of galaxies (e.g., \citealp{Bernardi2010, Bernardi2013, Bernardi2017}). The determination of a well-defined boundary is critical for measuring the photometric mass tracers. However, the velocity dispersion measurement, particularly using fiber spectroscopy, only requires obtaining the spectrum at the central region of galaxies. Thus, the determination of stellar velocity dispersion is insensitive to the complicated photometric procedure. Second, measuring stellar velocity dispersion is less affected by the complicated assumptions on baryonic physics used for measuring other mass tracers. For example, stellar mass measurement has a systematic variation depending on the assumption of the initial mass function (IMF) slope, metallicity, and star formation history (e.g., \citealp{Conroy2009}). In contrast, these variations in underlying stellar populations have little impact on the velocity dispersion measurements (e.g., \citealp{Knabel2025}). 

One key advantage of using stellar velocity dispersion is that it provides an independent probe of complex baryonic feedback processes, particularly feedback from active galactic nuclei (AGNs). AGN feedback was introduced to explain the observed deficit of massive galaxies relative to predictions from DM-only models of galaxy-mass distributions (e.g., \citealp{Ciotti1997, Silk1998, Menci2002, Benson2003, Binney2004, Silk2005, Springel2005a, Li2009}). In massive galaxies, AGN feedback plays a critical role by heating or expelling cold gas through the activity of central supermassive black holes, thereby suppressing star formation (e.g., \citealp{Schawinski2007, Combes2017}) and limiting the growth of stellar mass. The AGN feedback reaches out to $\sim 10$ kpc and may affect the stellar mass growth within the region (e.g., \citealp{Zubovas2016, Harrison2024}). This feedback mechanism has been incorporated into cosmological simulations resulting in broad consistency with observed galaxy luminosity and stellar mass distributions \citep{Schaye2015, Weigel2016, Pillepich2018}. However, the detailed physical processes driving AGN feedback remain poorly understood (e.g., \citealp{Wagner2012, Li2018}), and its influence on stellar kinematics, particularly on the distribution of stellar velocity dispersions, has not been thoroughly explored.

In this work, we investigated the stellar velocity dispersion of galaxies in cosmological numerical simulations with various AGN feedback models implemented. We particularly used the stellar mass to velocity dispersion relation and the stellar velocity dispersion function (VDF) to examine the impact of AGN feedback on the stellar velocity dispersion. The VDF is a statistical tool that displays the distribution of galaxy stellar velocity dispersions. In observations, several studies measure the VDFs for quiescent galaxies in field \citep{Sheth2003, Choi2007, Chae2010, Montero-Dorta2017, Sohn2017b} or in galaxy clusters \citep{Sohn2017a, Sohn2022} based on large-scale spectroscopic surveys. To provide the theoretical framework for these observed VDF measurements, \citet{Sohn2024b} found that the cluster VDFs are significantly shifted toward the lower velocity dispersion compared to the observed VDFs. They suggest that understanding the origin of the difference in the observed and simulated VDFs can provide an important test for galaxy formation models implemented in cosmological simulations. We specifically examined whether variations in AGN feedback can account for the apparent offset between the observed and simulated VDFs. We also investigated how the shapes of the VDFs change across different AGN feedback models and compared them with the observed VDF to identify which models best reproduce the data. 

Our investigation of the role of AGN feedback in galaxy kinematics is distinctive from previous examinations based on luminosity or stellar mass distribution because the simulations were not calibrated to match the stellar velocity dispersion distributions. Thus, the investigation of the stellar velocity dispersion distribution provides an independent test bed for AGN feedback models based on the statistical distribution of galaxy properties. \citet{Choi2007} explored the impact of AGN feedback on various galaxy properties including the velocity dispersions based on numerical simulations. Our study extends their work based on a much larger sample provided by cosmological simulations with various AGN feedback models. 

In Section \ref{sec:data}, we introduce the EAGLE simulations with different AGN feedback models we used. In Section \ref{sec:vdf}, we derive the velocity dispersion functions of galaxies in EAGLE simulations. Based on the comparison between VDFs, we investigated the impact of AGN feedback in the velocity and stellar mass distributions of galaxies in Section \ref{sec:discussion}. We conclude our results in Section \ref{sec:conclusion}. Throughout the paper, we adopt the Planck cosmological parameters for EAGLE-run cosmology consistent with \citet{PlanckCollaboration2014}: $\Omega_{m}$ = 0.307, $\Omega_{\Lambda}$ = 0.693, $\Omega_{b}$ = 0.048, and $h$ = 0.6777 (= $H_{0}$/(100 km s$^{-1}$ Mpc$^{-1}$)).

\section{Data} \label{sec:data}

\subsection{EAGLE} \label{sec:EAGLE}

We used Evolution and Assembly of GaLaxies and their Environments (EAGLE) cosmological simulations \citep{Crain2015, Schaye2015, McAlpine2016} to study the velocity dispersions of galaxies. EAGLE includes simulations covering various box sizes with various mass resolutions and baryonic physics models. Here, we used the intermediate-resolution simulations with the initial baryonic particle mass $m_{\rm g} = 1.81 \times 10^{6}~\Mdot$ and DM particle mass of $m_{\rm dm} = 9.70 \times 10^{6}~\Mdot$, covering a 50 Mpc box. We only used the $z = 0$ snapshot in this work. Because our primary goal is to compare the shapes of the velocity dispersion functions across different AGN feedback models, a single redshift snapshot is sufficient. Moreover, we compare the velocity dispersions of simulated galaxies with those of observed galaxies at $z < 0.2$. \citet{Zahid2016} showed that the observed scaling relations between stellar velocity dispersion and stellar mass exhibit little to no redshift evolution. Therefore, using simulated galaxies from the $z = 0$ snapshot provides a reasonable basis for comparison with observed galaxies at $z < 0.2$.

We used simulations covering a comoving volume of 50 Mpc$^{3}$ from the suite of EAGLE simulations. We chose this volume because it includes runs implemented with various AGN feedback models \citep{Schaye2015}. RefL0050N0752 (hereafter, EAGLE-50) is the simulation that uses the standard subgrid physics and numerical techniques of the EAGLE simulation (see more details in \citealp{Schaye2015}). There are four other simulations with different AGN feedback, including C15AGNdT9L0050N0752 (hereafter, EAGLE-eAGN-dT), ViscHiL0050N0752 (hereafter, EAGLE-eAGN-Visc), C15AGNdT8L0050N0752 (hereafter, EAGLE-wAGN), and NoAGNL0050N0752 (hereafter, EAGLE-NoAGN).

There are two parameters that determine the AGN feedback models in EAGLE simulations: $\Delta T_{\rm AGN}$ and $C_{\rm visc}$. The five simulations we used here are based on AGN feedback models with various combinations of $\Delta T_{\rm AGN}$ and $C_{\rm visc}$. $\Delta T_{\rm AGN}$ is the temperature increase of the gas during AGN feedback. A larger $\Delta T_{\rm AGN}$ indicates a more powerful feedback event. $C_{\rm visc}$ is the parameter related to the viscosity of a subgrid accretion disk \citep{Crain2015, Rosas-Guevara2015}. The lower $C_{\rm visc}$ indicates a higher disk viscosity by definition. In a simulation with a lower $C_{\rm visc}$ model, the black-hole growth and the resulting AGN feedback are delayed.

EAGLE-50 is the reference model with the standard AGN feedback model with $\Delta T_{\rm AGN} = 10^{8.5}$ K and $C_{\rm visc} = 2\pi$. EAGLE-eAGN-dT and EAGLE-eAGN-visc are simulations with enhanced AGN feedback models but different parameterizations. EAGLE-eAGN-dT assumes a more energetic AGN feedback with $\Delta T_{\rm AGN} = 10^{9}$ K, while EAGLE-eAGN-Visc assumes a lower $C_{\rm visc} = 2\pi \times 10^{-2}$ indicating a higher mass accretion rate of black holes resulting in a faster onset of the AGN feedback. EAGLE-wAGN employs a weaker AGN feedback with $\Delta T_{\rm AGN} = 10^{8}$ K compared to the standard EAGLE AGN feedback model. EAGLE-NoAGN assumes an extreme case where no black holes exist. Table \ref{tab:EAGLE} summarizes $\Delta T_{\rm AGN}$ and $C_{\rm visc}$ for these five simulations. 

\begin{table*}
\caption{Summary of EAGLE simulations used in this study.}
\centering
\label{tab:EAGLE}
\begin{tabular}{llcc}
\hline \hline
ID & Simulation\tablefootmark{a} & $\Delta T_{\rm AGN} \ [K] \tablefootmark{b}$ & $C_{\rm visc} \tablefootmark{c}$ \\
\hline
EAGLE-50        & RefL0050N0752        & $10^{8.5}$ & $2\pi$                \\
EAGLE-eAGN-dT   & C15AGNdT9L0050N0752  & $10^{9}$   & $2\pi$                \\
EAGLE-eAGN-Visc & ViscHiL0050N0752     & $10^{8.5}$ & $2\pi \times 10^{-2}$ \\
EAGLE-wAGN      & C15AGNdT8L0050N0752  & $10^{8}$   & $2\pi$                \\
EAGLE-NoAGN     & NoAGNL0050N0752      & --         & --                    \\
\hline
\end{tabular}
\tablefoot{
\tablefoottext{a}{EAGLE reference ID}
\tablefoottext{b}{$\Delta T_{\rm AGN}$ is the temperature increment of the gas during AGN feedback.}
\tablefoottext{c}{$C_{\rm visc}$ is a subgrid parameter related to the viscosity of the accretion disk of a subhalo.}}
\end{table*}

\begin{figure}
\centering
\includegraphics[width=0.8\linewidth]{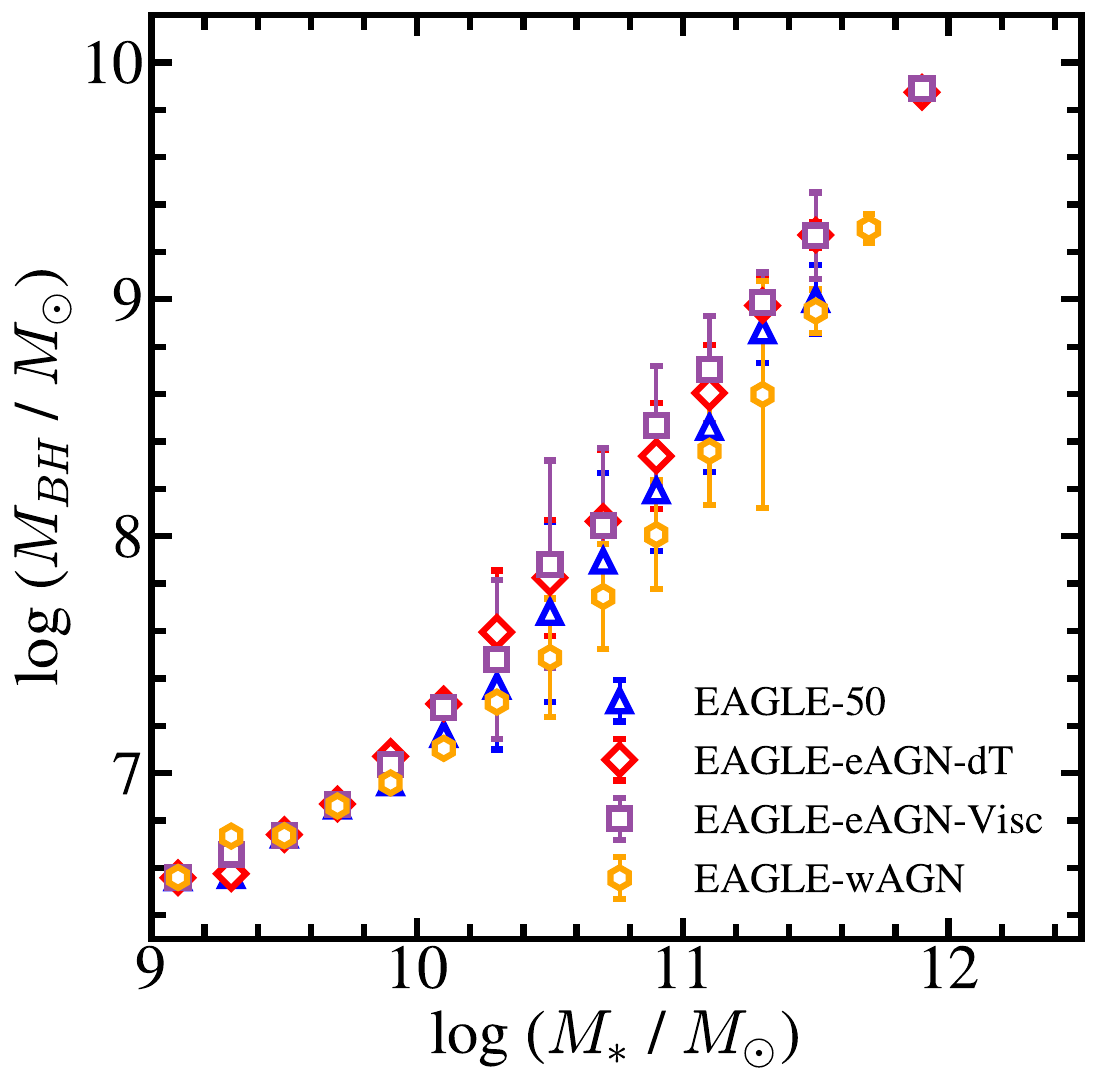}
\caption{Median black-hole mass as a function of stellar mass of subhalos in various EAGLE simulations at $z = 0$. Blue triangles, red diamonds, magenta squares, and orange hexagons show the relations from EAGLE-50, EAGLE-eAGN-dT, EAGLE-eAGN-Visc, and EAGLE-wAGN, respectively. At $11.6 < \log (M_{*}/ \Mdot) < 11.8$, EAGLE-50, EAGLE-eAGN-dT, EAGLE-eAGN-Visc do not contain any subhalos in the mass bin; there is no subhalo with $\log (M_{*}/ \Mdot) > 11.9$ in EAGLE-wAGN. Error bars show the 1$\sigma$ scatter in black-hole mass within each stellar mass bin.}
\label{fig:bhmass}
\end{figure}

Figure \ref{fig:bhmass} displays the mass of black holes ($M_{\rm BH}$) as a function of the stellar mass of their host subhalos ($M_{*}$) in four EAGLE simulations (except for EAGLE-NoAGN). The black-hole mass ($M_{\rm BH}$) is directly proportional to the feedback energy injected into the galaxies \citep{Goubert2024}. In general, $M_{\rm BH}$ increases as a function of $M_{*}$. The simulations with enhanced AGN feedback models (i.e., EAGLE-eAGN-dT and EAGLE-eAGN-Visc) show a steeper $M_{*} - M_{\rm BH}$ relation at $\log (M_{*}/ \Mdot) > 10$. In contrast, the black-hole mass in EAGLE-wAGN is lower than those in the other three simulations at fixed stellar mass.

\subsection{Identifying pressure-dominated subhalos in EAGLE}

We used subhalos in EAGLE simulations identified based on the SUBFIND algorithm \citep{Springel2001, Dolag2009}. We select subhalos with stellar masses larger than $10^{9}~\Mdot$, roughly corresponding to the absolute magnitude limits of dense spectroscopic surveys for the low redshift at $0.03 \leq z \leq 0.10$ \citep{Sohn2017b, Sohn2024a}. EAGLE simulations covering a 50 Mpc$^{3}$ volume that we used typically contain $\sim 1700$ subhalos.

We focused on galaxies with pressure-dominated kinematics to study the central stellar velocity dispersions. In the case of rotation-dominated galaxies, the ordered rotation may significantly affect the velocity dispersion measurements (e.g., \citealp{vanUitert2013}). We identified pressure-dominated galaxies based on the kinematics of stellar particles of subhalos. In observations, identifying pressure- or rotation-dominated galaxies is not straightforward without spatially resolved spectroscopy. Thus, to facilitate the comparison with the observational sampling of quiescent galaxies, the identification of quiescent galaxies based on a low specific star formation rate (e.g., \citealp{Zahid2016, Sohn2024a}) in simulations is often used. 

Unlike previous works, we used the kinematic properties to identify pressure-dominated galaxies because we only compared the simulated properties. Following \citet{Dubois2021}, we separated the pressure- and rotation-supported galaxies based on $v / \sigma = 0.5$, where $v$ and $\sigma$ indicate the rotational velocity and the velocity dispersion of stellar particles. 

We derived the rotational velocity ($v$) based on the angular momentum of stellar particles. To compute $v$, we first calculated the total angular momentum vector of a galaxy ($\textbf{J}_{\rm tot}$) based on the relative positions ($\textbf{x}_{i}$) and relative velocities ($\textbf{v}_{i}$) of stellar particles with respect to the center of the galaxy:
\begin{equation}
\textbf{J}_{\rm tot} = \sum_{i} m_{i} (\textbf{x}_{i} \times \textbf{v}_{i}).
\end{equation}
Then, we derived the principal axes ($\hat{\textbf{J}}_{\rm tot}$) of the angular momentum: 
\begin{equation}
\hat{\textbf{J}}_{\rm tot} = \textbf{J}_{\rm tot} / |\textbf{J}_{\rm tot}|.
\end{equation}
Here, $\hat{\textbf{J}}_{\rm tot}$ corresponds to the galaxy spin axis. Once the galaxy spin axis is defined, we then computed the angular-momentum vector of each particle:
\begin{equation}
\textbf{j}_{i} = \textbf{x}_{i} \times \textbf{v}_{i}.
\end{equation}
The angular momentum ($\textbf{j}_{z,i}$) and galaxy spin axis (perpendicular to rotational plane) of each stellar particle is defined as
\begin{equation}
{j}_{z, i} = \textbf{j}_{i} \cdot \hat{ \textbf{J}}_{\rm tot}.
\label{eq:principal}
\end{equation}
Additionally, we computed the projected distance on the rotational plane ($R_{i}$) of individual stellar particles from the galaxy spin axis. Based on these quantities, we finally computed rotational velocity of galaxies based on stellar particles within the stellar half-mass radius: 
\begin{equation}
v = <j_{z,i} / R_{i}>
\end{equation}
(see also \citealp{Sales2010, Sales2012}). 

We also computed the stellar velocity dispersion along with the galaxy spin axes we derived from Equation \ref{eq:principal}. The 1D stellar velocity dispersion is defined as 
\begin{equation}
\sigma = \sqrt{(\sigma_{r}^{2} + \sigma_{\theta}^{2} + \sigma_{z}^{2})/3}.
\end{equation}
We used the stellar particles within the half-mass radius the same as in the rotational velocity. Finally, we computed the $v/\sigma$ of all galaxies based on the definition from \citet{Dubois2021}.

\begin{figure*}
\centering
\includegraphics[width=1\linewidth]{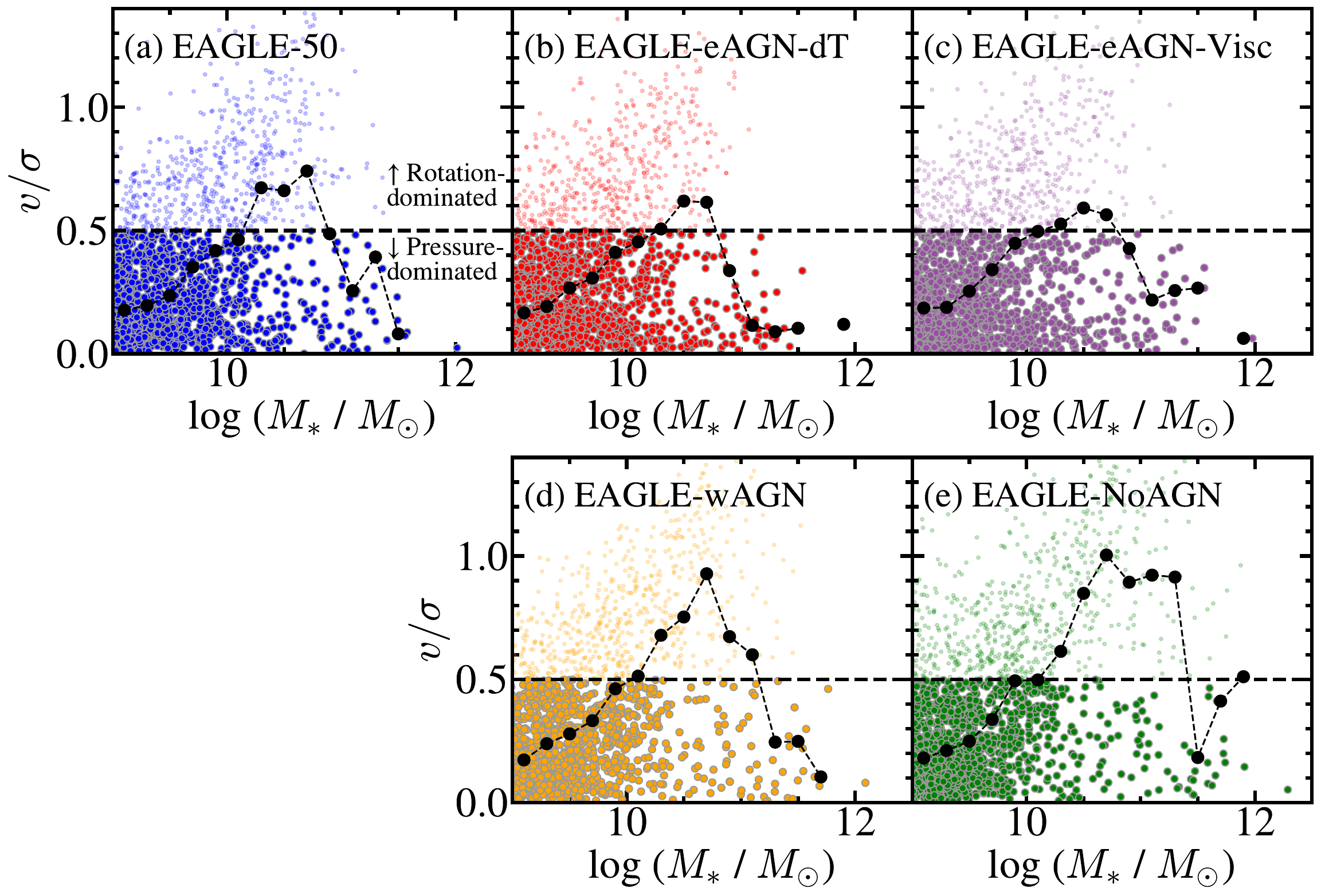}
\caption{$v/\sigma$ ratio as function of stellar mass for subhalos in (a) EAGLE-50, (b) EAGLE-eAGN-dT, (c) EAGLE-eAGN-Visc, (d) EAGLE-wAGN, and (e) EAGLE-NoAGN. Black circles display the median $v/\sigma$ as a function of stellar mass for subhalos in each simulation. Horizontal dashed lines indicate $v/\sigma = 0.5$, where the rotation-dominated ($v/\sigma \geq 0.5$) and pressure-dominated ($v/\sigma < 0.5$) galaxies are separated.} 
\label{fig:mvsigma}
\end{figure*}

Figure \ref{fig:mvsigma} shows the $v / \sigma$ of galaxy subhalos as a function of their stellar mass in the five EAGLE simulations. Black circles in each panel indicate the median $v / \sigma$ in each stellar mass bin. Horizontal lines mark the boundary separating pressure- and rotation-dominated galaxies.

The $v / \sigma - M_{*}$ relations vary depending on the AGN feedback models, particularly at $M_{*} > 10^{10}~\Mdot$. Compared to EAGLE-50 with the normal AGN feedback model, the simulations with enhanced AGN feedback models tend to have a smaller number of rotation-dominated galaxies. In contrast, the simulations with weak or no AGN feedback contain a much larger number of rotation-dominated galaxies. The variation of the $v / \sigma$ relation depending on the AGN feedback model is consistent with the comparison in other simulations: Horizon-AGN versus Horizon-NoAGN (see Figure 3 in \citealp{Dubois2016}). The differences in the fractions of rotation- and pressure-dominated galaxies suggest that AGN feedback plays a role in determining galaxy kinematics. However, the variation in the  $v / \sigma - M_{*}$ relation across the five EAGLE simulations is less than $\sim 10 \%$ in the entire sample of $M_{*} > 10^{9}~\Mdot$. Indeed, the $v / \sigma - M_{*}$ relation changes little at $M_{*} < 10^{10}~\Mdot$. Table \ref{tab:separation} summarizes the number of pressure- and rotation-dominated systems in the five simulations.

\begin{table}
\caption{Number of total, pressure- and rotation-dominated galaxies in EAGLE simulations.}
\centering
\label{tab:separation}
\begin{tabular}{lcc}
\hline \hline
Simulation                & Galaxy selection    & N ($M_{*} > 10^{9}~\Mdot$) \\
\hline
\multirow{3}{*}{EAGLE-50}        & Total               & 1754 \\
                                 & $v/\sigma < 0.5$    & 1194 \\
                                 & $v/\sigma \geq 0.5$ &  560 \\
\hline
\multirow{3}{*}{EAGLE-eAGN-dT}   & Total               & 1723 \\
                                 & $v/\sigma < 0.5$    & 1227 \\
                                 & $v/\sigma \geq 0.5$ &  496 \\
\hline
\multirow{3}{*}{EAGLE-eAGN-Visc} & Total               & 1714 \\
                                 & $v/\sigma < 0.5$    & 1197 \\
                                 & $v/\sigma \geq 0.5$ &  517 \\
\hline
\multirow{3}{*}{EAGLE-wAGN}      & Total               & 1707 \\
                                 & $v/\sigma < 0.5$    & 1110 \\
                                 & $v/\sigma \geq 0.5$ &  597 \\
\hline
\multirow{3}{*}{EAGLE-NoAGN}     & Total               & 1774 \\
                                 & $v/\sigma < 0.5$    & 1141 \\
                                 & $v/\sigma \geq 0.5$ &  633 \\
\hline
\end{tabular}
\end{table}

\subsection{stellar velocity dispersion} \label{sec:sigma}

We derived the stellar velocity dispersion ($\sigma_{*}$) of pressure-dominated galaxies in the simulations. We also note that the stellar velocity dispersion we derived here differs slightly from the velocity dispersion used to compute $v / \sigma$. Following \citet{Sohn2024a}, which derived $\sigma_{*}$ from IllustrisTNG (see also \citealp{Sohn2022, Sohn2024b}), we computed the velocity dispersion of member stellar particles of subhalos within a cylindrical volume that penetrates the center of subhalos. In particular, we computed the line-of-sight velocity dispersion along with the $z-$axis, mocking the velocity dispersion measurements based on fiber spectroscopy. We computed the stellar mass-weighted standard deviation as stellar velocity dispersions.

\citet{Sohn2024a} investigated the velocity dispersions based on various definitions in IllustrisTNG, including different viewing axes and measurement techniques. They showed that the 1D velocity dispersions measured along the $x-$, $y-$, and $z-$axes are generally consistent with each other. Furthermore, they derived velocity dispersions based on various techniques, including standard deviation, luminosity- or stellar mass-weighted standard deviation, and bi-weight measurements. All velocity dispersion measurements are consistent with each other. Similarly, the velocity dispersion measurements for EAGLE simulations are also insensitive to the various definitions of velocity dispersion. We thus used the stellar mass-weighted standard deviation of stellar particles, along with the $z-$axis as our velocity dispersion measurements (i.e., $\sigma_{*, z}$). 

\begin{figure}
\centering
\includegraphics[width=0.8\linewidth]{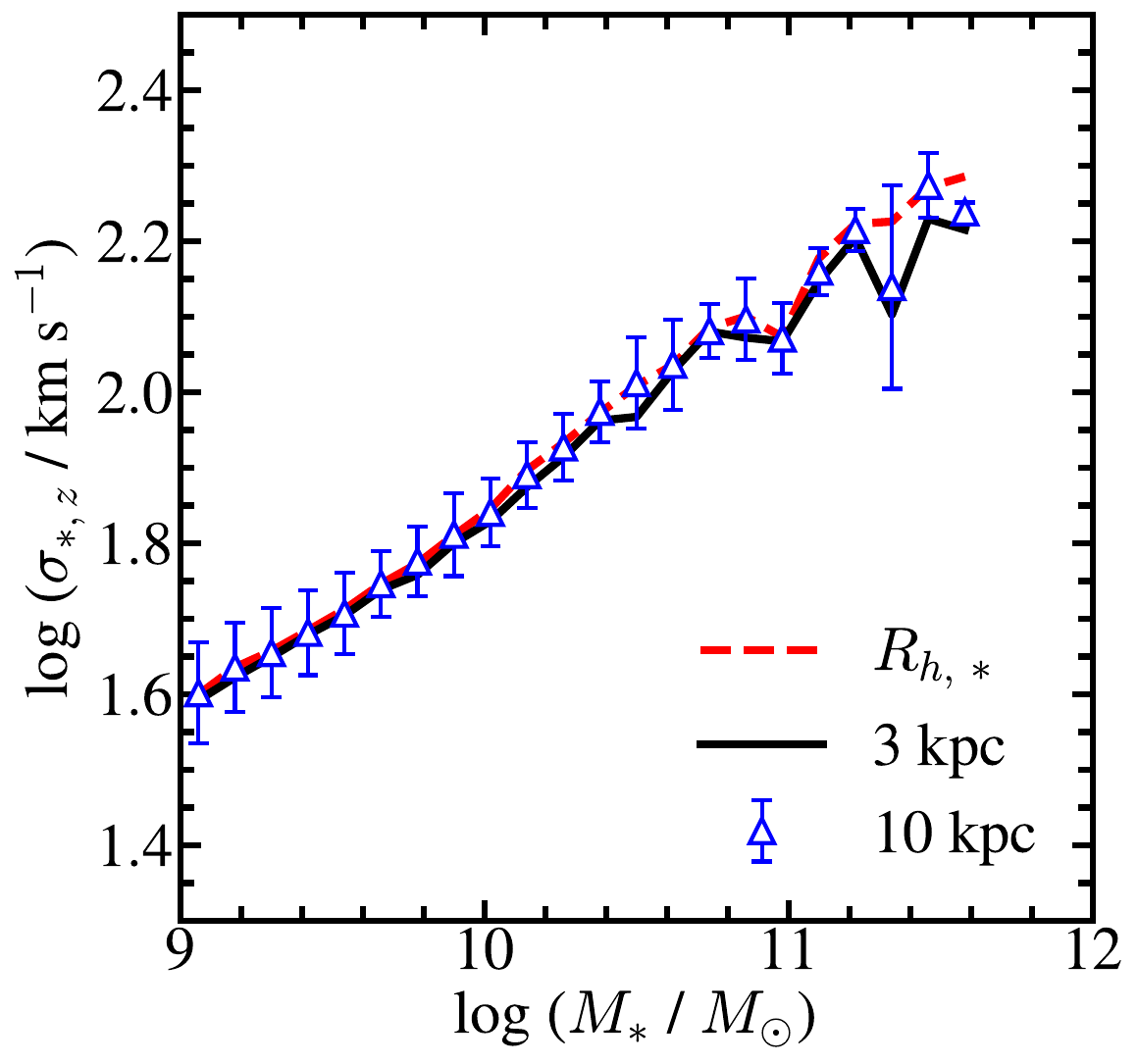}
\caption{Median line-of-sight velocity dispersions (along with $z-$axis) of pressure-dominated galaxies in EAGLE-50 measured within a cylindrical volume with an aperture of 10 kpc (blue triangles). The error bar indicates the 1$\sigma$ standard deviation in each stellar mass bin. For comparison, the median velocity dispersions are measured within 3 kpc (solid black line), and the half-mass radius is defined by the stellar particles (dashed red line).}
\label{fig:msigmastar}
\end{figure}

We adopted a fixed 10 kpc aperture to measure stellar velocity dispersions. Using a fixed aperture facilitates comparison with observational measurements, which are typically derived from fiber spectroscopy with fixed angular sizes. \citet{Sohn2024b} and \citet{Sohn2024a} used a 3 kpc fixed aperture to study the stellar velocity dispersions of subhalos in the IllustrisTNG-300 simulation. Here, we employed a slightly larger aperture of 10 kpc, which is sufficiently large to mitigate the effects of gravitational softening in the simulations (i.e., $\sim 0.7$ kpc for EAGLE simulations; \citealp{Schaye2015}). Figure \ref{fig:msigmastar} shows the relationship between stellar velocity dispersion measured within 10 kpc ($\sigma_{*, 10~\mathrm{kpc}}$) and stellar mass ($M_{*}$) for subhalos in the EAGLE-50 simulation. For comparison, we also show the same relation using velocity dispersions measured within 3 kpc and $R_{h,*}$ apertures (the solid and dashed lines in Figure \ref{fig:msigmastar}). The resulting $M_{*} - \sigma_{*}$ relations are generally consistent with one another, suggesting that the choice of aperture does not significantly affect the subsequent analysis. Hereafter, we refer to $\sigma_{*, z, 10~\rm kpc}$ as $\sigma_{*}$.

\section{Results} \label{sec:vdf}

\subsection{The $M_{*} - \sigma_{*}$ Relations}

\begin{figure*}
\centering
\includegraphics[width=1\linewidth]{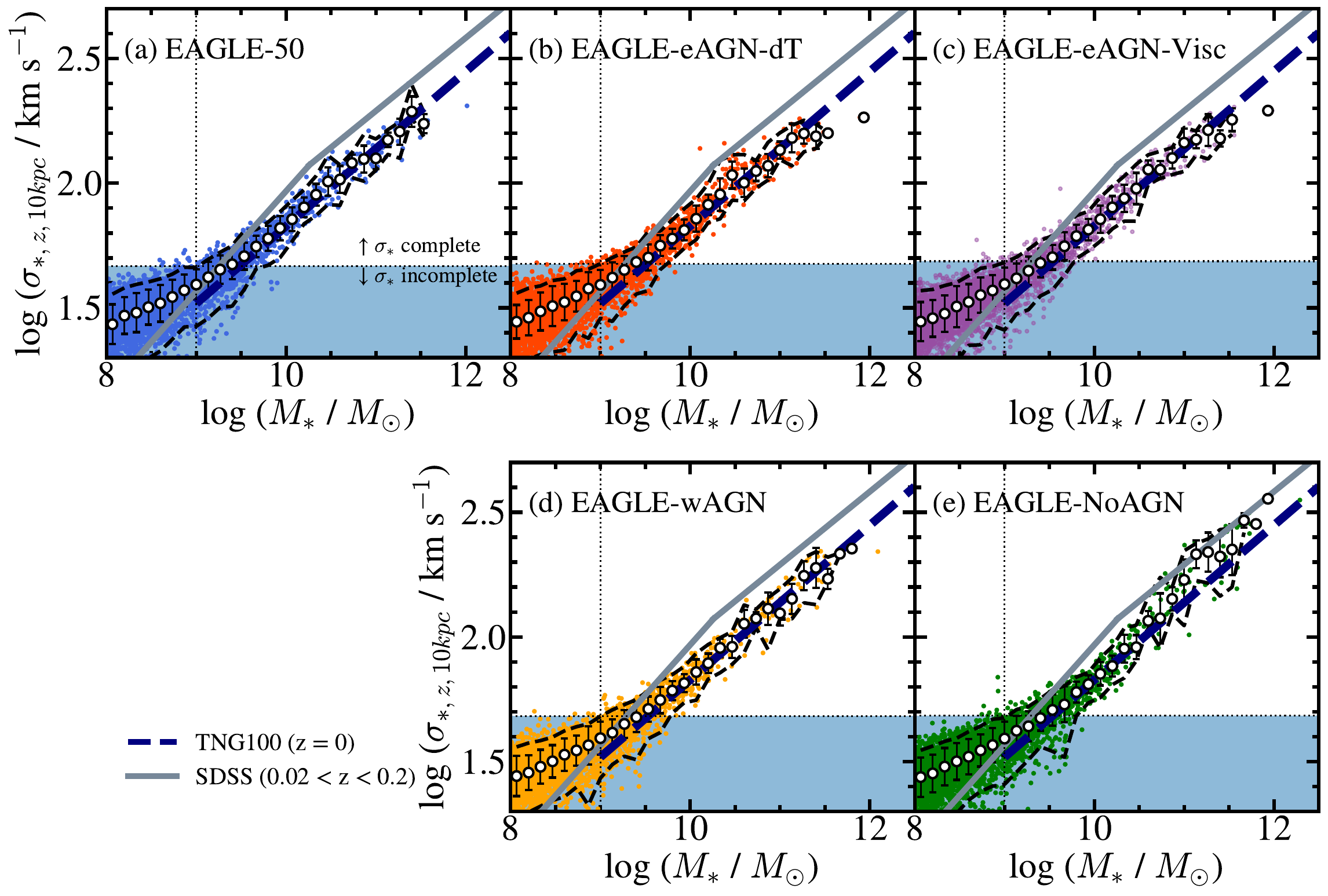}
\caption{Stellar mass to 1D stellar velocity dispersion measured within a cylindrical volume with a 10 kpc aperture along with $z-$axis ($\sigma_{*, z, 10 \rm ~kpc}$) relations of galaxy subhalos in (a) EAGLE-50, (b) EAGLE-eAGN-dT, (c) EAGLE-eAGN-Visc, (d) EAGLE-wAGN, and (e) EAGLE-NoAGN. White circles show the median $\sigma_{*, z, 10\rm ~kpc}$ at each stellar mass bin. Bold, dashed black lines show the central 95\% of the $\sigma_{*, z, 10\rm ~kpc}$ distributions. Vertical dotted lines indicate the stellar mass limit we applied. Horizontal dotted lines mark the velocity dispersion  limit where the stellar mass limit intersects with the 95\% upper limit of the velocity dispersion  distributions. Below this velocity dispersion  limit, our EAGLE galaxy samples are incomplete in terms of velocity dispersion. The subhalos in the shaded region are excluded from the $\sigma_{*}$ complete sample. The dashed blue lines indicate the $M_{*} - \sigma_{*}$ relation for TNG100 \citep{Sohn2024a}, and the solid gray lines show the observed $M_{*} - \sigma_{*}$ relation for SDSS quiescent subhalos in the local Universe \citep{Zahid2016}.}
\label{fig:sigmacom}
\end{figure*}

Figure \ref{fig:sigmacom} compares the $M_{*} - \sigma_{*}$ relations of five EAGLE simulations. In all five simulations, the stellar velocity dispersion of pressure-dominated galaxies is generally proportional to their stellar mass. The $M_{*} - \sigma_{*}$ relations derived from EAGLE simulations with AGN feedback (regardless of the strength) are consistent with each other. However, the $M_{*} - \sigma_{*}$ relation from the EAGLE-NoAGN differs from those from the other simulations; the EAGLE-NoAGN velocity dispersions are larger than those in other simulations at a given stellar mass, particularly at $M_{*} > 10^{10.5}$ M$_{\odot}$. 

The solid gray lines in Figure \ref{fig:sigmacom} display the observed $M_{*} - \sigma_{*}$ relation from \citet{Zahid2016}. The observed relation is derived from quiescent galaxies with D$_{n}$4000 larger than 1.5 selected from SDSS spectroscopic sample within $0.02 < z < 0.2$. The observed relation is based on $\sigma_{*}$ measured within a 3 kpc aperture. Because the simulated velocity dispersions we computed vary little within 3 kpc and 10 kpc apertures, we compared the observed relation without applying any aperture correction. The observed relation shows a broken power law with different slopes at $M_{*} = 10^{10.26}$ M$_{\odot}$ (see also \citealp{Nigoche-Netro2011, Cappellari2013a, Cappellari2016}). 

The $M_{*} - \sigma_{*}$ relation for EAGLE-NoAGN galaxies is much closer to the observed relation compared to the similar relations for galaxies in other simulations (Figure \ref{fig:sigmacom}). In particular, galaxies with $M_{*} > 10^{11}$ M$_{\odot}$ in EAGLE-NoAGN have velocity dispersions comparable with the observed galaxies. This apparent consistency clearly indicates that the AGN feedback has an impact on the velocity dispersion distributions of galaxies. We discuss the impact of the AGN feedback in Section \ref{sec:role}. 

We also compare the EAGLE $M_{*} - \sigma_{*}$ relations with the IllustrisTNG-100 (TNG100) simulation. IllustrisTNG simulations use the AGN feedback models implementing the thermal injection (quasar mode) and kinetic wind (kinetic mode) from the AGN \citep{Weinberger2018, Pillepich2018}, while EAGLE simulations only use the quasar mode \citep{Schaye2015}. We chose TNG100 because the mass resolution of TNG100 is comparable with the EAGLE-50 simulations we used. There are TNG50 simulations covering a similar volume with EAGLE-50 simulations, but their mass resolution is an order of magnitude better than the EAGLE-50 simulation. \citet{Sohn2024a} demonstrated that the $M_{*} - \sigma_{*}$ relations from TNG50 and TNG100 are offset; the stellar velocity dispersions for TNG50 galaxies are slightly larger than that for TNG100 galaxies at a given stellar mass. 

We used the TNG100 $M_{*} - \sigma_{*, z, 3~\rm kpc}$ relation derived by \citet{Sohn2024a}. The TNG100 relation is derived from quiescent galaxies with specific star formation rates lower than $2 \times 10^{-11}$ yr$^{-1}$ at $z = 0$. Because \citet{Sohn2024a} showed that the velocity dispersions measured within 3 kpc and 10 kpc are comparable, we used the $M_{*} - \sigma_{*, z, 3~\rm kpc}$ relation from \citet{Sohn2024a} without any aperture correction. 

In Figure \ref{fig:sigmacom}, the dashed navy line shows the TNG100 $M_{*} - \sigma_{*, z, 3~\rm kpc}$ relation. The TNG100 relation describes the distribution of EAGLE-50 quiescent galaxies very well, except for galaxies in the EAGLE-NoAGN simulation. The relations derived from both simulations differ significantly from the observed relation, as \citet{Sohn2024a} already pointed out. In other words, the $M_{*} - \sigma_{*}$ relations derived from EAGLE and TNG100 simulations with AGN feedback are all coincident at $M_{*} > 10^{10.5}~\Mdot$.

\subsection{The velocity dispersion  complete samples}

The VDF measures the number of galaxies within bins of velocity dispersion. Deriving the VDF in both observations and simulations can be affected by systematic incompleteness introduced by sample selection. For example, \citet{Sohn2017b} showed that constructing the VDF from a volume-limited sample leads to incompleteness in velocity dispersion, primarily due to the large scatter between velocity dispersion and galaxy magnitude. Similarly, deriving VDFs from simulations also requires care in the construction of velocity dispersion -complete samples \citep{Sohn2024b}.

We started constructing the velocity dispersion -complete samples from the stellar mass-limited samples (i.e., $M_{*} > 10^{9}~\Mdot$) in five EAGLE simulations. Figure \ref{fig:sigmacom} shows that stellar velocity dispersions are generally proportional to the stellar mass, but with significant scatters. Because of this scatter, the determination of the velocity dispersion  completeness limit of the stellar mass-limited sample is not straightforward. 

Following \citet{Sohn2024b}, we determined the velocity dispersion  completeness limit of the EAGLE stellar mass-limited samples (see also \citealp{Sohn2017b}). We first derived the velocity dispersion  boundaries (dashed black lines in Figure \ref{fig:sigmacom}), where 95\% of galaxies in each stellar mass bin are included. We then computed the intercept between the upper 95\% boundary and the stellar mass limit (i.e., $M_{*} = 10^{9}~\Mdot$). This intercept (horizontal lines in Figure \ref{fig:sigmacom}) indicates the velocity dispersion  limit of the stellar mass-limited sample. In other words, the galaxies with a stellar mass lower than the stellar mass limit may have larger velocity dispersion than the velocity dispersion  limit. Thus, constructing the VDF based on the stellar mass-limited sample may cause the incompleteness of the VDF below the velocity dispersion  completeness limit. 

The velocity dispersion  completeness limits are $\log \sigma_{*} \approx 1.7$ across all five simulations. The velocity dispersion  completeness limit for EAGLE simulations is consistent with the completeness limit derived from TNG stellar mass-limited sample with $\log (M_{*}/\Mdot) > 9$ ($\log \sigma_{*} \approx 1.7$; \citealp{Sohn2024b}). We thus compared the EAGLE and TNG VDFs without concern for differences in completeness. The completeness limits for both EAGLE and TNG are slightly lower than the observed limit ($\log \sigma_{*} \approx 1.84$) derived from SDSS magnitude- and volume-limited samples \citep{Sheth2003, Choi2007, Sohn2017b}.

The final velocity dispersion  complete samples for five EAGLE simulations contain 550 -- 750 pressure-dominated galaxies. We used these velocity dispersion -complete samples for our VDF measurements in Section \ref{sec:vdf_result}. Table \ref{tab:number} summarizes the number of pressure-dominated galaxy samples in EAGLE simulations that we used. 

\begin{table*}
\caption{Number of pressure-dominated galaxies in EAGLE simulations.}
\centering
\label{tab:number}
\begin{tabular}{llcc}
\hline \hline
Simulation      &  Galaxy selection  & N$_{M_{*}{\rm -limited}}$ $\tablefootmark{a}$ & N$_{\sigma_{*}{\rm-limited}}$ $\tablefootmark{b}$  \\
\hline
\multirow{2}{*}{EAGLE-50}        & $v/\sigma < 0.5$                     & 1194 & 728 \\
                                 & sSFR $< 2 \times 10^{-11}$ yr$^{-1}$ & 554  & 321 \\
\hline
\multirow{2}{*}{EAGLE-eAGN-dT}   & $v/\sigma < 0.5$                     & 1227 & 728 \\
                                 & sSFR $< 2 \times 10^{-11}$ yr$^{-1}$ & 635  & 437 \\
\hline
\multirow{2}{*}{EAGLE-eAGN-Visc} & $v/\sigma < 0.5$                     & 1197 & 647 \\
                                 & sSFR $< 2 \times 10^{-11}$ yr$^{-1}$ & 736  & 476 \\
\hline
\multirow{2}{*}{EAGLE-wAGN}      & $v/\sigma < 0.5$                     & 1110 & 596 \\
                                 & sSFR $< 2 \times 10^{-11}$ yr$^{-1}$ & 490  & 283 \\
\hline
\multirow{2}{*}{EAGLE-NoAGN}     & $v/\sigma < 0.5$                     & 1141 & 563 \\
                                 & sSFR $< 2 \times 10^{-11}$ yr$^{-1}$ & 416  & 200 \\
\hline
\end{tabular}
\tablefoot{
\tablefoottext{a}{N$_{M_{*}{\rm-limited}}$ is the number of pressure-dominated galaxies with $M_{*} > 10^{9}~\Mdot$.}
\tablefoottext{b}{N$_{\sigma_{*}{\rm-limited}}$ is the number of pressure-dominated galaxies in the $\sigma_{*}$ -complete sample defined in each EAGLE simulation (see Section \ref{sec:vdf}).}
}
\end{table*}

\subsection{The VDFs in EAGLE simulations} \label{sec:vdf_result}

\begin{figure*}
\centering
\includegraphics[width=1\linewidth]{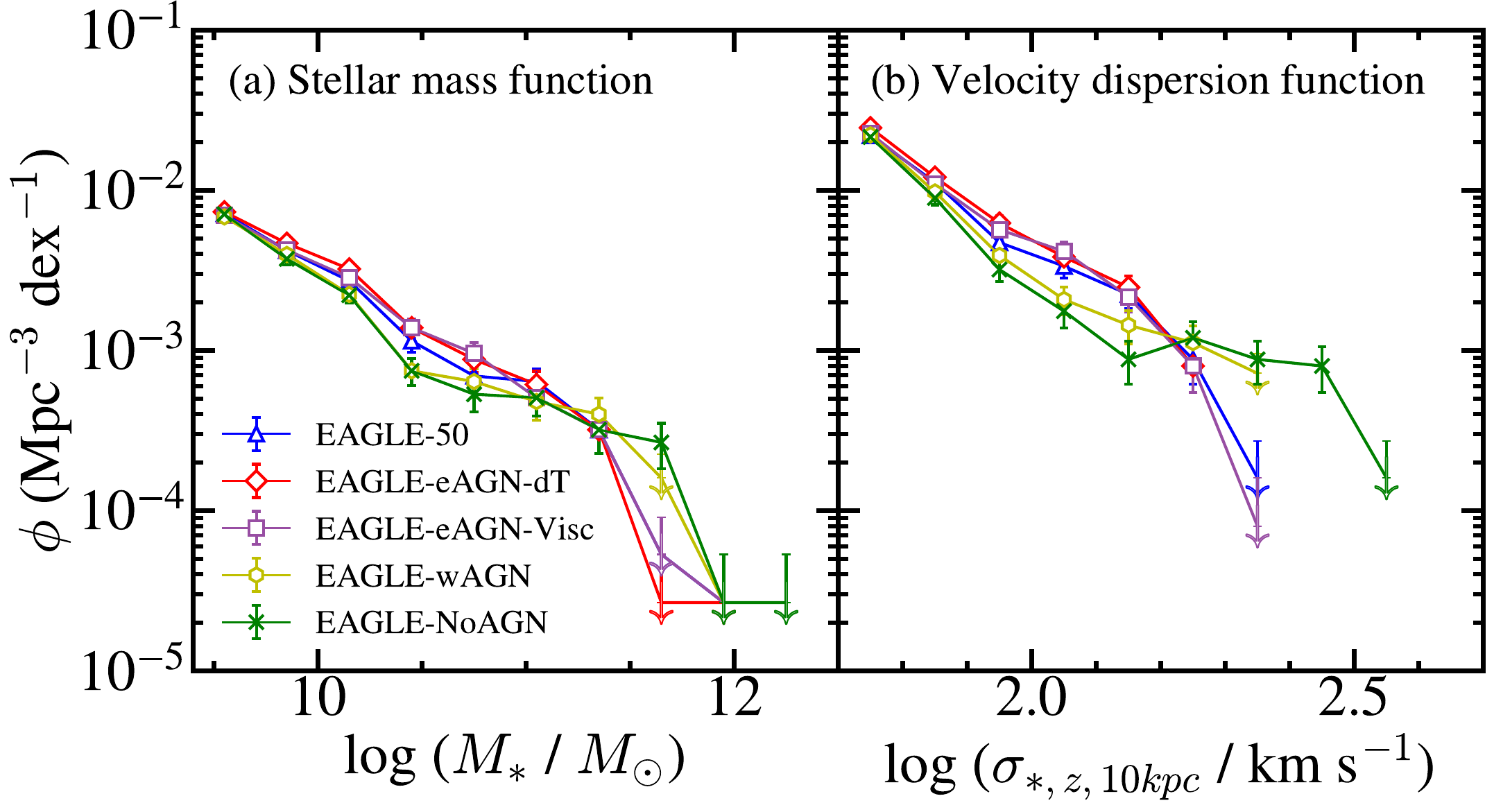}
\caption{(a) Stellar mass functions and (b) VDFs of pressure-dominated galaxies in five EAGLE simulations. Green crosses show the relations from EAGLE-NoAGN. The other symbols are the same as in Figure \ref{fig:bhmass}. Error bars indicate Poisson errors. For bins with fewer than ten galaxies, we only display the upper values, which are marked by arrows.}
\label{fig:smfvdf}
\end{figure*}

Figure \ref{fig:smfvdf} (a) shows the stellar mass functions (SMFs) of pressure-dominated galaxies in the $\sigma_{*}$ -complete samples for the five EAGLE simulations. We normalized the SMF by the total three-dimensional volume of the simulations. Blue triangles, red diamonds, magenta squares, orange hexagons, and green crosses represent the SMFs for EAGLE-50, EAGLE-eAGN-dT, EAGLE-eAGN-Visc, EAGLE-wAGN, and EAGLE-NoAGN, respectively. Error bars represent Poisson uncertainties. We note that some high stellar mass or stellar velocity dispersion bins contain fewer than ten galaxies. For these bins, we only show the upper limit, defined as the upper end of the $1\sigma$ confidence interval, and omit the lower error bar.

The SMFs from the simulations with normal or enhanced AGN feedback models show consistent stellar mass distributions. The simulations with enhanced AGN feedback models contain slightly fewer massive galaxies, as expected, but the difference is insignificant. The SMFs derived from simulations with weak and no AGN feedback are flatter at $10.5 < \log (M_{*} / \Mdot) < 12$ than the SMFs from simulations with AGN feedback. Massive galaxies with $\log (M_{*} /\Mdot) > 11$ are more abundant in simulations with no or weak AGN feedback. This comparison indicates that the AGN feedback in EAGLE simulations indeed suppresses the formation of massive galaxies. 

In Figure \ref{fig:smfvdf} (b), we compare the VDFs measured based on $\sigma_{*}$ complete samples for the five EAGLE simulations. Similar to the SMFs, the VDFs from EAGLE-50, EAGLE-eAGN-dT, and EAGLE-eAGN-Visc are consistent with each other. The VDFs from EAGLE-wAGN and EAGLE-NoAGN differ from other VDFs. These simulations with weaker AGN feedback contain significantly more subhalos with $\log \sigma_{*} > 2.3$. The number of galaxies with $2.0 < \log \sigma_{*} < 2.3$ in EAGLE-wAGN and EAGLE-NoAGN is relatively lower than in other simulations. We discuss the origin of the differences in VDFs depending on the feedback model in Section \ref{sec:role}.

\subsection{Comparison between simulated \& observed VDFs} \label{sec:simobs}

We next compare the observed and simulated VDFs of pressure-dominated galaxies. We used the observed VDF from \citet{Sohn2017b} constructed from the quiescent galaxies in the SDSS spectroscopy. \citet{Sohn2017b} selected quiescent galaxies with $D_{n}4000 > 1.5$ within the $0.03 \leq z \leq 0.10$ redshift range from the SDSS main galaxy sample. They obtained the stellar velocity dispersion measured within a fiducial 3 kpc aperture. \citet{Sohn2017b} demonstrated the importance of building the velocity dispersion -complete sample for measuring the VDF; they showed that the SDSS VDF measured based on the velocity dispersion -complete sample is flatter at $\log \sigma_{*} < 2.1$ compared to the SDSS VDF measured based on the magnitude- or volume-limited samples \citep{Sheth2003, Choi2007, Chae2010}. We therefore adopted the field VDF from \citet{Sohn2017b}, derived from a velocity dispersion  complete SDSS sample, as it provides the most complete measurement among existing VDF estimates.

\begin{figure*}
\centering
\includegraphics[width=1\linewidth]{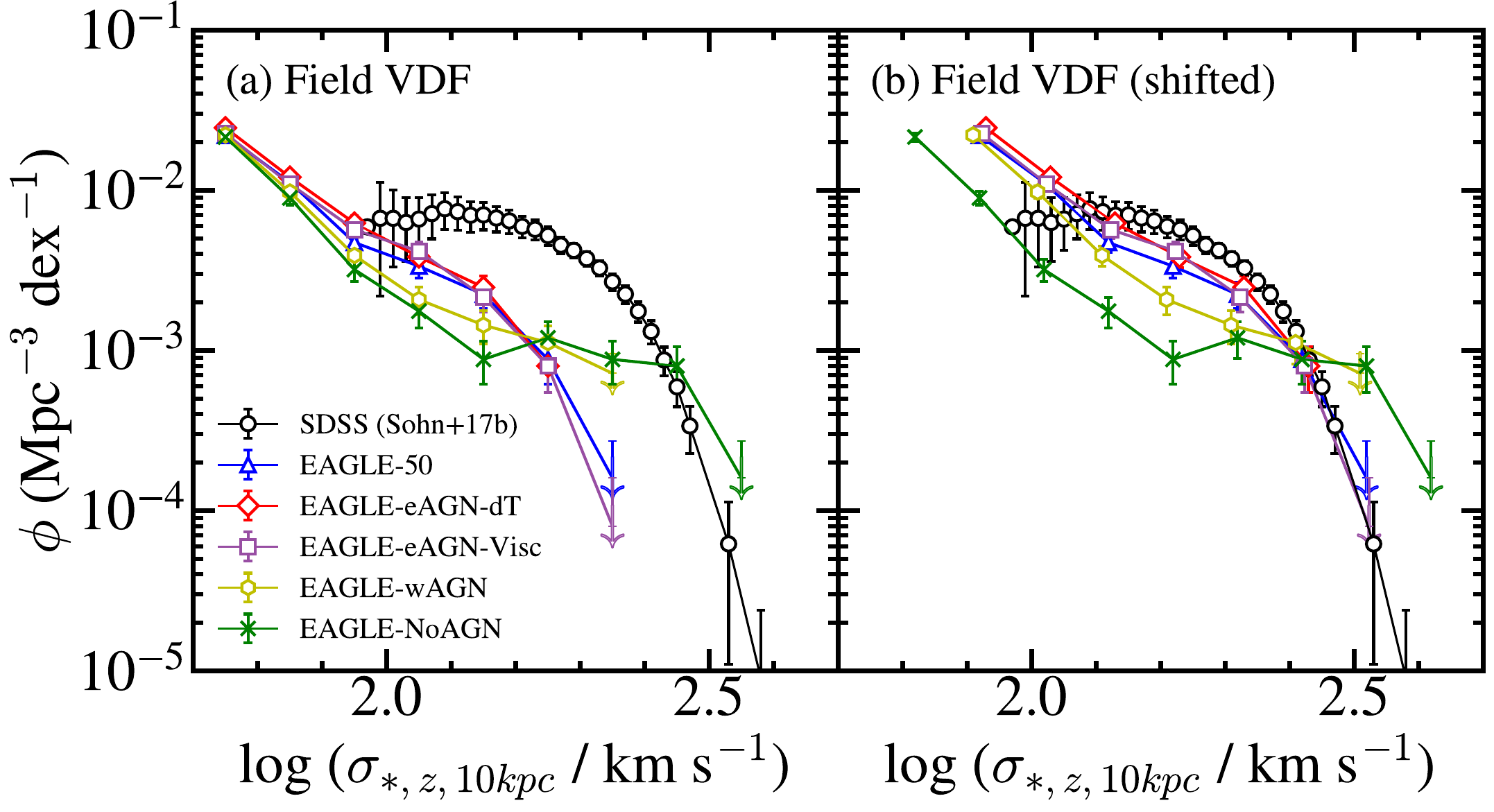}
\caption{(a) Comparison between observed and simulated VDFs. Black circles show the observed VDF derived from SDSS field galaxies at $0.03 \leq z \leq 0.10$ \citep{Sohn2017b}. Colored symbols display the simulated VDFs from five EAGLE simulations. Symbols are identical from Figure \ref{fig:smfvdf}. (b) Same as (a), but the simulated VDFs are shifted by the $\sigma_{*}$ offset determined based on the $M_{*} - \sigma_{*}$ relations. Error bars on the solid lines indicate Poisson errors. For bins with fewer than ten galaxies, we only display the upper values, marked by arrows.}
\label{fig:vdfshifted}
\end{figure*}

Figure \ref{fig:vdfshifted} (a) directly compares the observed SDSS VDF (the black line and circles) with the VDFs measured from five EAGLE simulations. All five simulated VDFs show significant offsets compared to the observed VDF. The VDF from EAGLE-NoAGN has a comparable value at $\log \sigma_{*} > 2.4$, but it also deviates from the observed VDF at lower velocity dispersions. The offset between the observed and EAGLE VDFs corresponds to $\Delta \log \sigma_{*} \sim 0.15 - 0.18$ dex, consistent with the offset in the $M_{*} - \sigma_{*}$ relations. The significant offset between observed and EAGLE VDFs is consistent with the result from the comparison between the observed and TNG VDFs \citep{Sohn2024b}. 

Figure \ref{fig:vdfshifted} (b) shows the simulated VDFs after shifting in the $\sigma_{*}$ direction, horizontally. To determine the amount of shift, we calculate the mean velocity dispersion  offset between the observed and simulated $M_{*} - \sigma_{*}$ relations (see Figure \ref{fig:sigmacom}) for more than 30 galaxies within $11 < \log (M_{*} / \Mdot) < 12$. We applied this mean offset when shifting the VDFs to preserve their overall shapes. Interestingly, the shapes of the simulated VDFs from the normal and enhanced AGN feedback are consistent with the shape of the observed VDF. In particular, the slope of the simulated VDFs at $\log \sigma_{*} > 2.4$ is consistent with the observed VDF. On the contrary, the VDFs derived from weak or no AGN feedback show a much flatter slope at a high-velocity dispersion  end. 

\section{Discussion} \label{sec:discussion}

We derived the VDFs of pressure-dominated galaxies in five EAGLE simulations with various AGN feedback models implemented. The shapes of VDFs measured from these simulations vary; particularly, the VDF measured from EAGLE-NoAGN shows a much flatter slope, indicating the relative effective formation of high-velocity dispersion  galaxies. We also compared these simulated VDFs with the observed VDF. To understand the implications of the differences in VDF shapes, we first examine the impact of sample-galaxy selection in Section \ref{sec:vsigmassfr}. We then discuss the role of AGN feedback in our interpretation of the implication of differences in simulated VDFs in Section \ref{sec:role}.

\subsection{Galaxy-sample selection} \label{sec:vsigmassfr}

In Figure \ref{fig:vdfshifted}, we show that the EAGLE simulations predict a significantly lower number of galaxies in the $100 < \sigma (\rm km \ s^{-1}) < 250$ velocity dispersion  range than the observations. The offset between observed and simulated VDFs is consistent with previous results based on IllustrisTNG simulations \citep{Sohn2024b}. However, we used the different galaxy identification compared to that applied in both observed \citep{Sohn2017b} and IllustisTNG galaxy samples \citep{Sohn2024b}. In this work, we identified pressure-dominated galaxies using dynamical criteria (i.e., $v / \sigma$), which are not straightforward to apply to observed galaxy samples. In contrast, \citet{Sohn2017b} identified quiescent galaxies with D$_{n}4000 > 1.5$, and \citet{Sohn2024b} selected galaxies with a specific star formation rate lower than $2 \times 10^{-11}$ yr$^{-1}$ in IllustrisTNG300 as quiescent galaxies for their VDF measurements. We thus investigated if the different identification of galaxy samples affects the derivation of VDFs. 

\begin{figure*}
\centering
\includegraphics[width=1\linewidth]{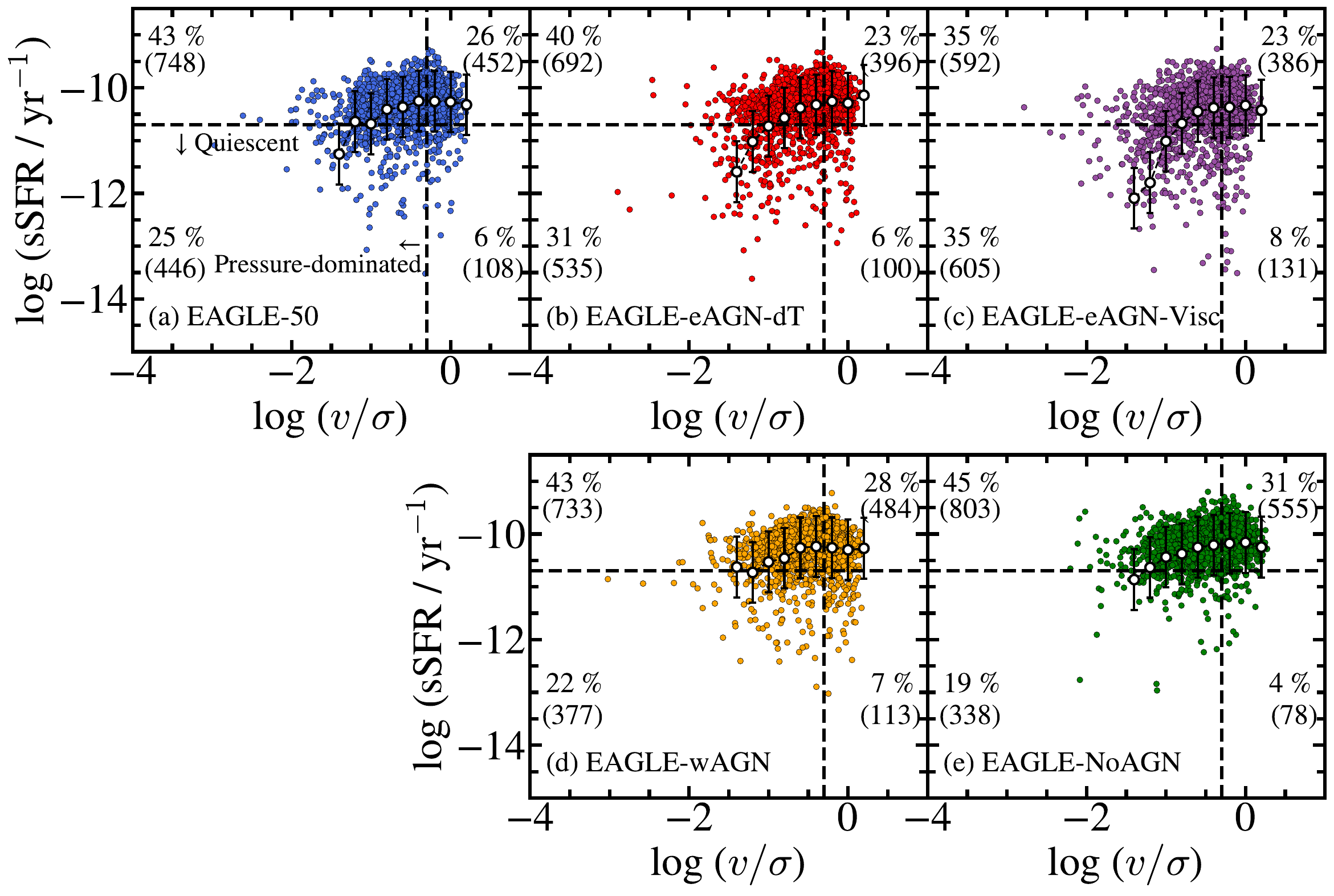}
\caption{Specific star formation rate as a function of $v/\sigma$ for all subhalos in five EAGLE simulations. Vertical lines show the selection boundary between pressure-dominated and rotation-dominated galaxies (i.e., $v/\sigma = 0.5$). Horizontal lines show another boundary that separates quiescent and star-forming galaxies (i.e., sSFR$ = 2 \times 10^{-11}$ yr$^{-1}$). The numbers indicate the fraction of galaxies in each area among the entire sample with $\log (M_{*}/\Mdot) > 9$. White circles show the median sSFR at each $v/\sigma$ bin. The error bar indicates the 1$\sigma$ standard deviation of the galaxies in each $v/\sigma$ bin.}
\label{fig:ssfrvsigma}
\end{figure*}

Figure \ref{fig:ssfrvsigma} shows the specific star formation rate (sSFR) as a function of $v/\sigma$ for subhalos for $M_{*} > 10^{9}~\Mdot$ in the five EAGLE simulations. In Figure \ref{fig:ssfrvsigma}, galaxies below the horizontal line are classified as quiescent galaxies, which were identified using the same selection criteria as in \citet{Sohn2024b} (hereafter the quiescent galaxy sample). Galaxies to the left of the vertical line are pressure-dominated systems (hereafter the pressure-dominated sample). The overlap between the quiescent and pressure-dominated samples is surprisingly small. Among quiescent galaxies, $\sim 75 - 85~\%$ are pressure-dominated galaxies, whereas only $\sim 30 - 50~\%$ of pressure-dominated galaxies are quiescent galaxies. This difference in sample selection influences the shape of the resulting VDFs.

\begin{figure}
\centering
\includegraphics[width=0.8\linewidth]{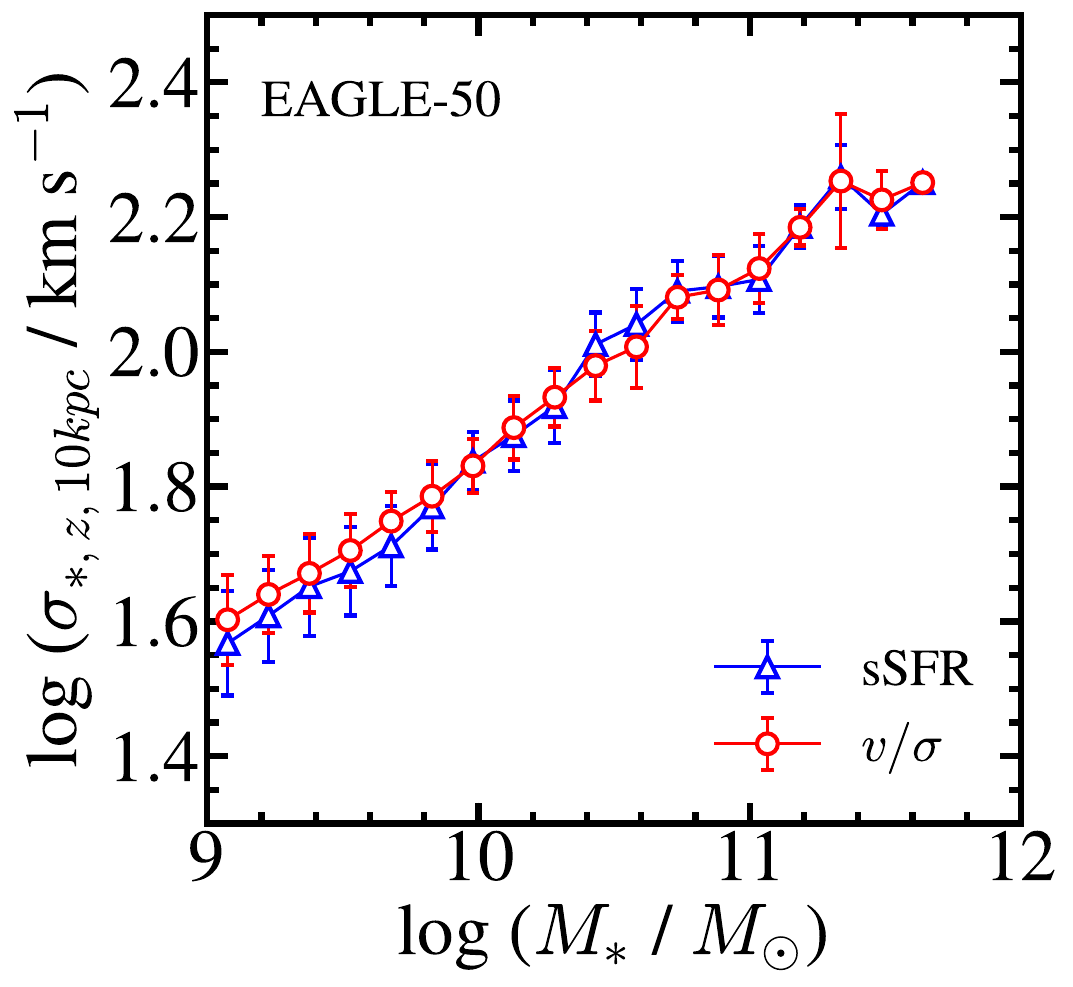}
\caption{Median $M_{*} - \sigma_{*}$ relation for EAGLE-50. Red circles and blue triangles show the relations derived from pressure-dominated galaxies with $v/\sigma < 0.5$ and the quiescent galaxies with sSFR $< 2\times 10^{-11}$ yr$^{-1}$, respectively. The error bars indicate the $1\sigma$ standard deviations.}
\label{fig:sigmaz10comp}
\end{figure}

Figure \ref{fig:sigmaz10comp} compares the $M_{*} - \sigma_{*}$ relations for the pressure-dominated sample ($v/\sigma < 0.5$, red circles) and the quiescent sample (sSFR $< 2 \times 10^{-11}$ yr$^{-1}$, blue triangles) in the EAGLE-50 simulation. The median $M_{*} - \sigma_{*}$ relations for the two samples are consistent over the mass range we explored. In other words, the stellar velocity dispersion does not change significantly at fixed stellar mass, regardless of how galaxies are identified. We also note that the pressure-dominated and quiescent samples in other EAGLE simulations are also consistent without any offset. 

\begin{figure}
\centering
\includegraphics[width=0.9\linewidth]{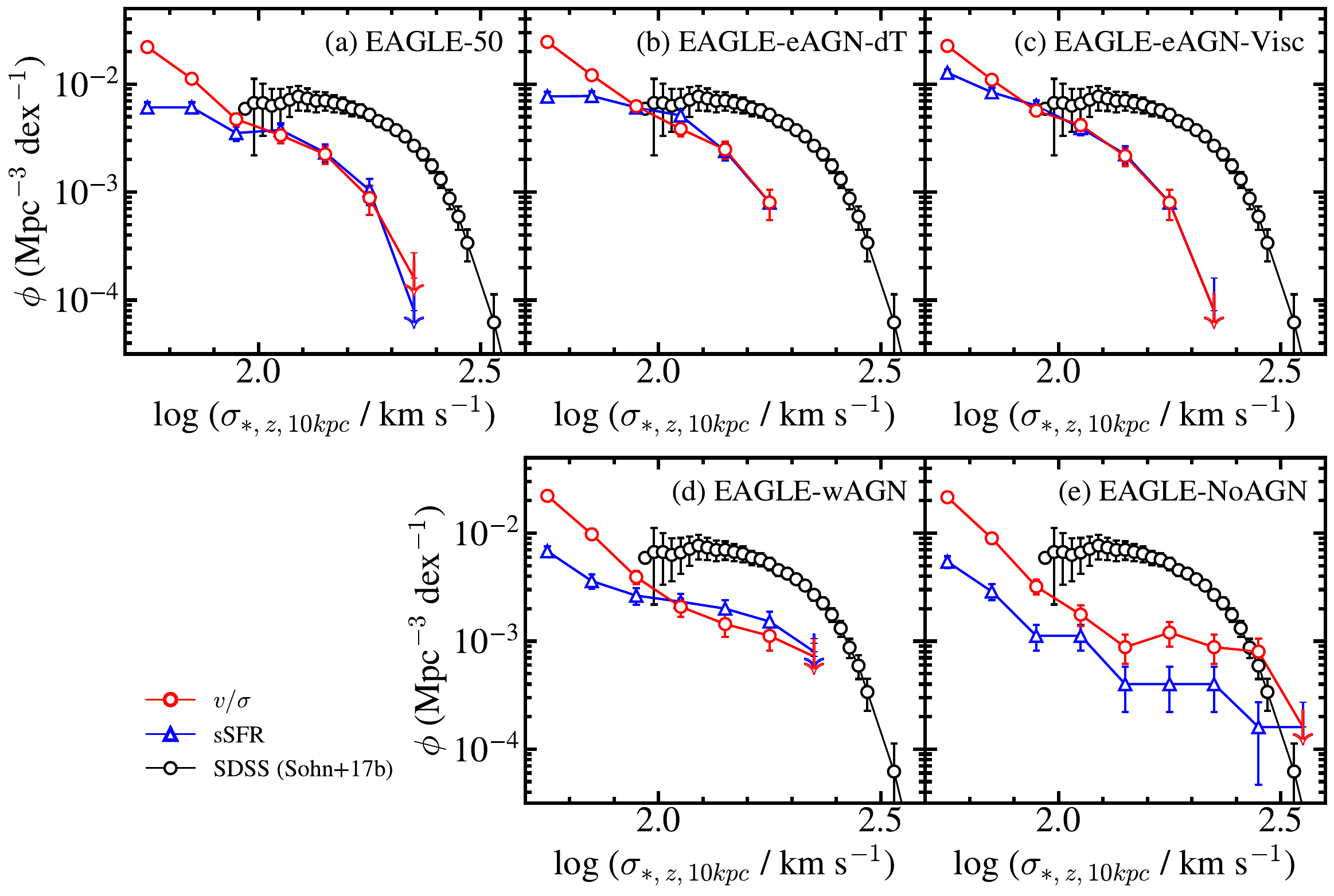}
\caption{Comparison between VDFs constructed from galaxy samples identified using $v / \sigma$-based and sSFR-based selections in five EAGLE simulations. Black circles show the observed VDF derived from SDSS field galaxies at $0.03 \leq z \leq 0.10$ \citep{Sohn2017b}. Symbols are identical to those of Figure \ref{fig:sigmaz10comp}. Error bars on the solid lines indicate Poisson errors. We only display the upper values with arrows for bins including fewer than ten galaxies.}
\label{fig:vdfcomp}
\end{figure}

Finally, Figure \ref{fig:vdfcomp} compares the VDFs derived from two different samples: the quiescent galaxy sample (blue triangles) and the pressure-dominated sample (red circles). The VDFs are generally consistent at $\log \sigma_{*} > 2.0$, indicating that differences in galaxy classification have minimal impact on the VDF shape at high-velocity dispersions. However, at $\log \sigma_{*} < 2.0$, the VDF from the pressure-dominated sample exhibits a significantly steeper slope than that of the quiescent sample. This steepening is due to the inclusion of many pressure-dominated galaxies with ongoing star formation (i.e., galaxies in the upper left region of Figure \ref{fig:ssfrvsigma}). More importantly, our analysis indicates that differences in sample selection alone cannot account for the substantial discrepancy between the observed and simulated VDFs, regardless of the AGN feedback model adopted.

\subsection{The role of AGN feedback in velocity dispersion} \label{sec:role}

Our tests imply two important aspects of the VDFs derived from EAGLE simulations with various AGN feedback models. First, the VDFs derived from EAGLE simulations are significantly offset from the observed VDFs, similarly to the VDFs derived from other simulations including TNG100 and TNG300 \citep{Sohn2024b}. Second, the $M_{*} - \sigma_{*}$ relation and the VDF vary little among EAGLE simulations, except for the one with no AGN feedback model. The steeper $M_{*} - \sigma_{*}$ relation of the galaxies in the EAGLE-NoAGN simulation is intriguing because the detailed physical mechanism of the AGN feedback affecting the stellar kinematics is unclear. 

To understand the role of AGN feedback on the stellar kinematics, we examined the origin of the peculiarities in the $M_{*} - \sigma_{*}$ relation and the VDF from the EAGLE-NoAGN simulation. Unlike the $M_{*}-\sigma_{*}$ relation we show in Figure \ref{fig:sigmacom}, we investigated the stellar mass-to-$\sigma_{*}$ relation, but based on the $M_{*}$ measured within a 10 kpc aperture (i.e., $M_{*, z, 10~\rm kpc}$). 

\begin{figure}
\centering
\includegraphics[width=0.8\linewidth]{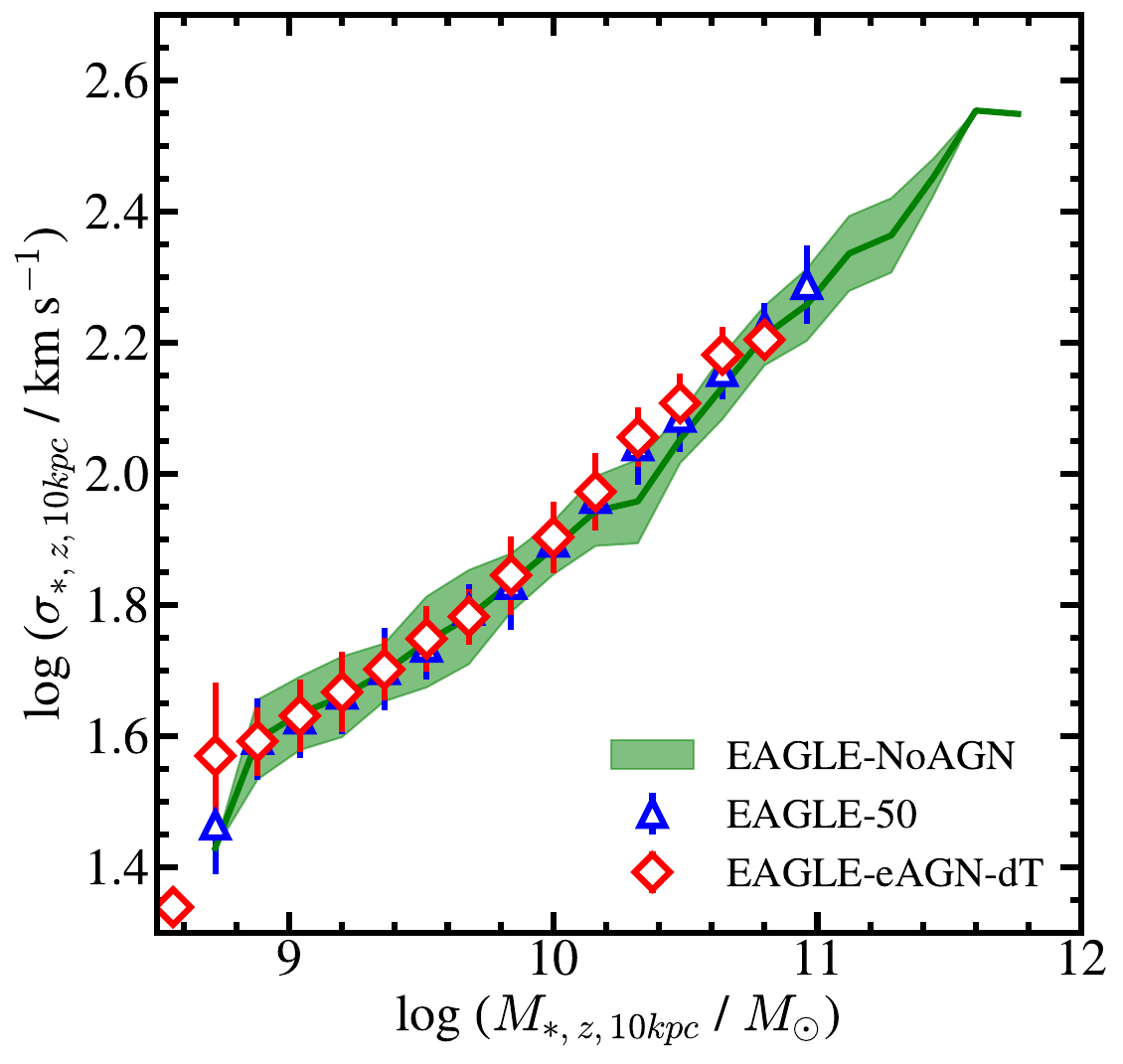}
\caption{Median stellar velocity dispersions measured within a cylindrical volume within a 10 kpc aperture as a function of the stellar mass measured within a 10 kpc aperture for pressure-dominated galaxies in various EAGLE simulations at $z = 0$. Blue triangles, red diamonds, and the solid green line show the relations from EAGLE-50, EAGLE-eAGN-dT, and EAGLE-NoAGN, respectively. The error bar and shaded region indicate the $1\sigma$ standard deviation of the galaxies in each stellar mass bin.}
\label{fig:apsigma}
\end{figure}

\begin{figure}
\centering
\includegraphics[width=0.8\linewidth]{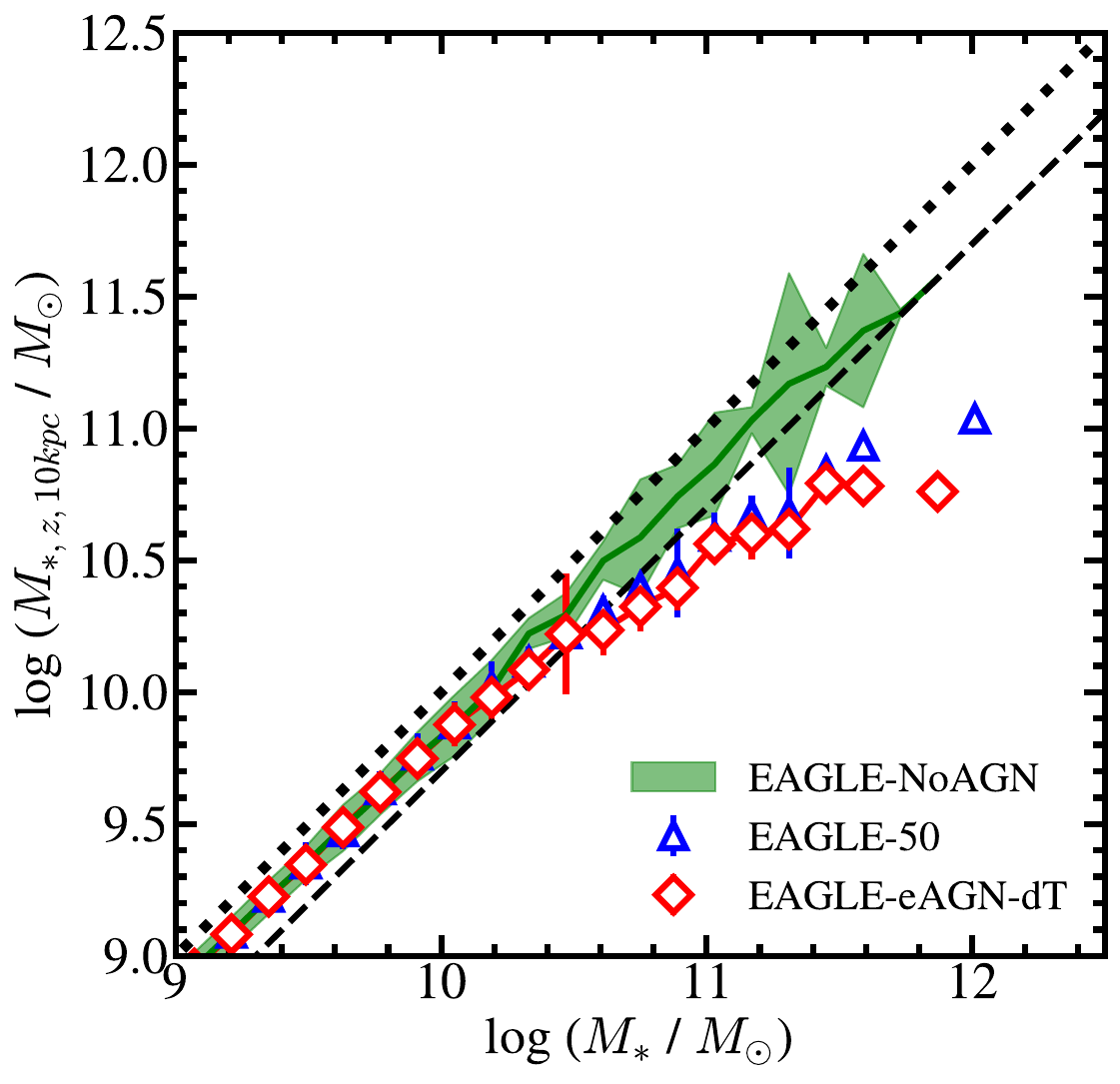}
\caption{Median stellar mass measured within cylindrical volume with 10 kpc aperture as function of total stellar mass of pressure-dominated galaxies in various EAGLE simulations. Symbols are the same as in Figure \ref{fig:apsigma}. The dashed line indicates the relations between total stellar mass and the stellar mass within half-mass radius. The dotted line indicates the one-to-one correspondence between the two stellar masses. The error bar and shaded region indicate the $1\sigma$ standard deviation of the galaxies in each stellar mass bin.}
\label{fig:apmass}
\end{figure}

In Figure \ref{fig:apsigma}, the $M_{*, 10~\rm kpc} - \sigma_{*, z, 10~\rm kpc}$ relations derived from three simulations (i.e., EAGLE-50, EAGLE-eAGN-dT, and EAGLE-NoAGN) do not differ from each other, unlike the $M_{*} - \sigma_{*, z, 10~\rm kpc}$ relations shown in Figure \ref{fig:sigmacom}. The discrepancy between the $M_{*} - \sigma_{*, z, 10~\rm kpc}$ and $M_{*, z, 10~\rm kpc} - \sigma_{*, z, 10~\rm kpc}$ relations indicates that 1) the stellar velocity dispersion is directly proportional to the stellar mass within the aperture where we measured the velocity dispersion; and 2) the stellar mass distribution within galaxies varies depending on the AGN feedback models (resulting in the different slopes in $M_{*} - \sigma_{*, z, 10~\rm kpc}$ relations). Figure \ref{fig:apmass} shows that the mass concentration of simulated galaxies varies depending on the adopted AGN feedback model. For $\log M_{*} < 10.5$, the galaxies in the EAGLE-NoAGN (the green line) continue to follow a proportional relation with $M_{*}$, whereas galaxies in the other EAGLE simulations show systematically lower $\log M_{*, z, 10~\rm kpc}$, reflecting reduced mass concentrations within 10 kpc due to AGN feedback.

Several physical mechanisms contribute to variations in the stellar mass distribution within galaxies (e.g., \citealp{Furlong2017, Choi2018}). First, galaxies naturally lose stellar mass as old stars through supernova explosions or during the asymptotic-giant-branch (AGB) phase. Second, stellar mass redistribution occurs through mergers or interactions between galaxies. These two processes likely operate to a similar extent regardless of the AGN feedback model employed.

The AGN feedback can change the stellar mass distribution by disturbing the gravitational potential (e.g., \citealp{Choi2018}). When AGN feedback expels gas from galaxy centers, it reduces the central gravitational potential, causing the inner regions of galaxies to "puff up." As a result, both the stellar and DM distributions expand, and the central potential becomes shallower (adiabatic expansion; e.g., \citealp{Hopkins2010, Fan2008, Fan2010}). This puff-up process generally increases the size of galaxies due to the migration of stars \citep{Furlong2017, Choi2018}. Galaxies in simulations with no AGN feedback are then compact compared to their counterparts in simulations with AGN feedback. Indeed, galaxies in EAGLE-NoAGN are generally more compact than those in other EAGLE simulations that include AGN feedback, at a fixed stellar mass. In the EAGLE-NoAGN simulation, galaxies with $M_{*} > 10^{11}~M_{\odot}$ have median half-mass radii of 3 -- 30 kpc, whereas galaxies with similar stellar masses in the other EAGLE simulations are much larger (10 -- 130 kpc), with median sizes about four times greater.

\begin{figure*}
\centering
\includegraphics[width=1\linewidth]{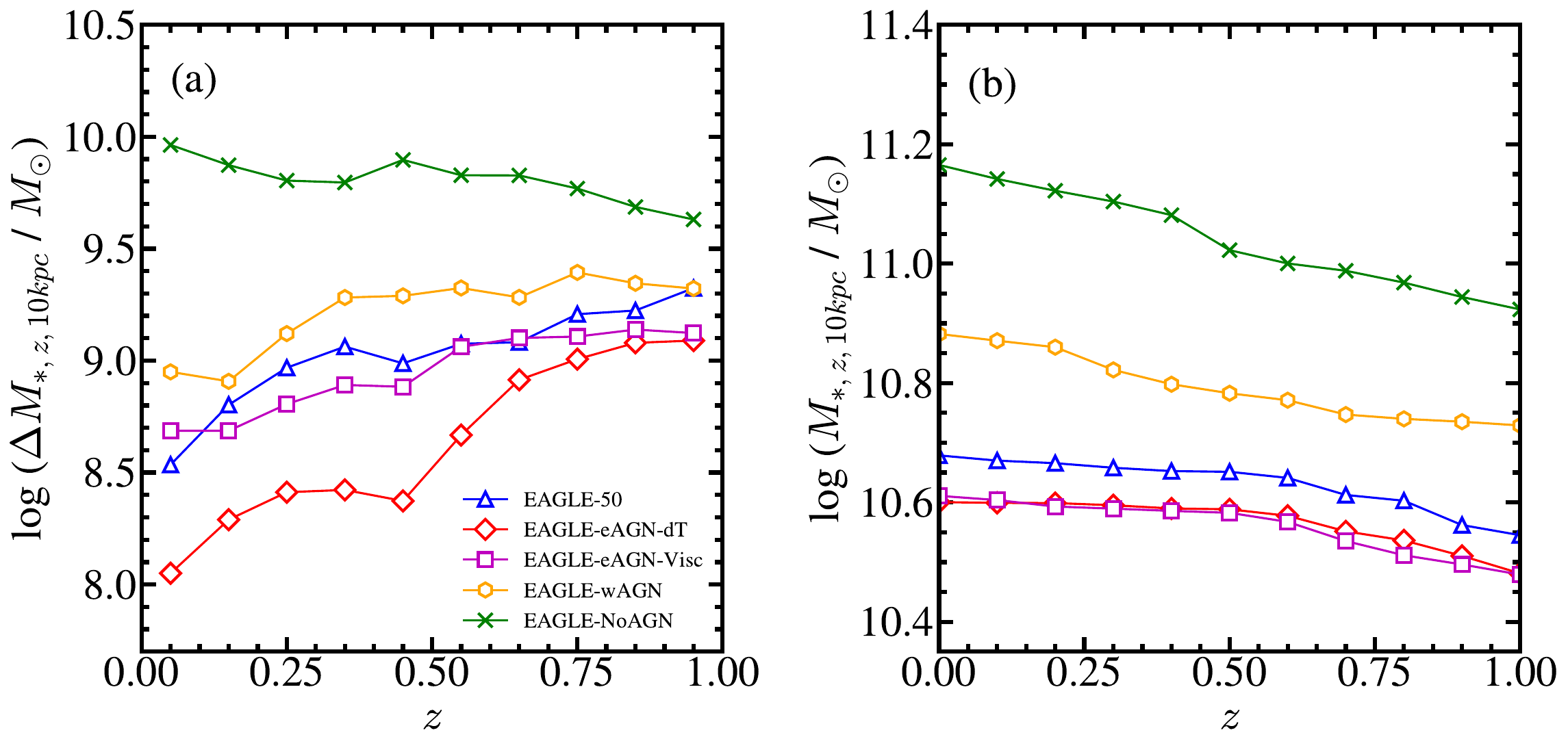}
\caption{(a) Median change of $M_{*, z, 10~\rm kpc}$ and (b) median of $M_{*, z, 10~\rm kpc}$ as a function of redshift in various EAGLE simulations at $z < 1$. The symbols are the same as in Figure \ref{fig:smfvdf}.}
\label{fig:delMstarz}
\end{figure*}

Figure \ref{fig:delMstarz} shows that the increase in stellar mass in the central region is strongly affected by AGN feedback. Figure \ref{fig:delMstarz} (a) shows the change of stellar mass within a 10 kpc aperture as a function of time. We plot the median variation of stellar mass of pressure-dominated galaxies at $z < 1$ in five EAGLE simulations. In general, galaxies in all EAGLE simulations show continuous mass growth at $< 10$ kpc. However, the median stellar mass growth rate varies significantly depending on the AGN feedback models. The mass-growth rate of galaxies simulated with the AGN feedback generally decreases as a function of time. In contrast, the mass-growth rate of galaxies in EAGLE-NoAGN exhibits a monotonic increase across the redshift range we explored.

Figure \ref{fig:delMstarz} (b) displays the resulting stellar mass growth at $z < 1$. Galaxies in EAGLE-NoAGN show a significant mass growth compared to galaxies in other EAGLE simulations. The significant growth of stellar mass within 10 kpc in EAGLE-NoAGN results in excess of $\sigma_{*, z, 10~\rm kpc}$s appeared in the $M_{*} - \sigma_{*, z, 10~\rm kpc}$ relation and the VDF. 

\begin{figure*}
\centering
\includegraphics[width=1\linewidth]{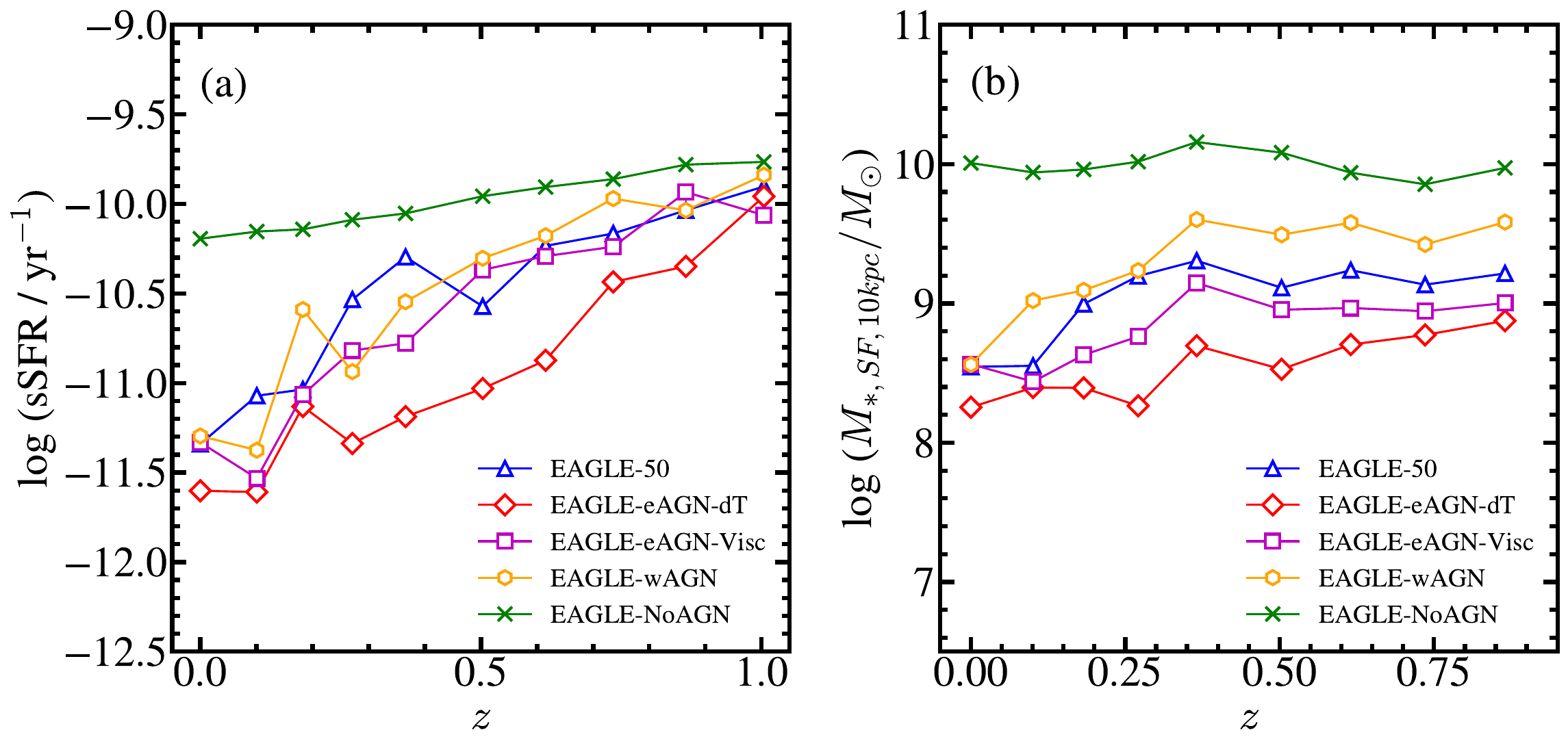}
\caption{(a) Median sSFR within spherical 10 kpc aperture in five EAGLE simulations as a function of redshift. (b) The median $M_{*, ~\rm SF, 10~\rm kpc}$, the amount of stellar mass within a 10 kpc spherical aperture increased by in situ star formation in various EAGLE simulations as a function of redshift at $z < 1$. The symbols are the same as in Figure \ref{fig:smfvdf}.}
\label{fig:sSFRz}
\end{figure*}

The most significant change in the stellar mass distribution due to the AGN feedback results from the variation in the star formation rate. Figure \ref{fig:sSFRz} (a) displays the median sSFR of pressure-dominated galaxies in EAGLE simulations. We note that the sSFR is measured within a spherical 10 kpc aperture for the galaxies instead of the cylindrical aperture as we examined for Figure \ref{fig:delMstarz}. The median sSFR of galaxies in EAGLE-NoAGN is higher than that of galaxies in other EAGLE simulations. Overall, the sSFR of galaxies decreases over time, but the sSFR of galaxies in EAGLE-NoAGN decreases relatively slowly compared to that of galaxies in other EAGLE simulations. 

Figure \ref{fig:sSFRz} (b) compares the amount of stellar mass within a 10 kpc spherical aperture increased by in situ star formation (i.e., $M_{*,~\rm SF, 10~kpc}$) at each redshift. To compute the in situ stellar mass increment, we simply assumed a linear variation of the star formation rate of galaxies between each redshift; then we computed the mass of newly formed stars in between the snapshots. The median $M_{*,~\rm SF, 10~\rm kpc}$ of galaxies in EAGLE-NoAGN remains constant at $z < 1$, while the median $M_{*,~\rm SF, 10~\rm kpc}$ of galaxies in other EAGLE simulations continuously decrease over time. The decrease in $M_{*,~\rm SF, 10~\rm kpc}$ observed in most EAGLE simulations is clearly a result of the reduced availability of cool gas due to AGN feedback. 

Additionally, we examined the ex situ stellar mass growth within a 10 kpc aperture. In the four EAGLE simulations that include AGN feedback, the ex situ mass growth is typically an order of magnitude larger than the in situ mass growth. In contrast, galaxies in the EAGLE-NoAGN simulation show comparable in situ and ex situ mass growth at $z < 1$. This indicates that the absence of AGN feedback leads to substantial stellar mass growth from internal star formation in the central region that is comparable to that driven by mergers or accretion.

Figure \ref{fig:delMstarz} and Figure \ref{fig:sSFRz} illustrate the different stellar mass evolution and evolution of stellar mass distribution of galaxies in EAGLE simulations with various AGN feedback models. In general, galaxies in all EAGLE simulations show stellar mass growth at $z < 1$ in the central region. The growth rate varies depending on the implemented AGN feedback model; the galaxies in EAGLE-NoAGN show much significant mass growth in the central region. When the AGN feedback is implemented, the central black hole activity suppresses star formation in the central region by heating the gas and expulsion of gas. As a result, the stellar mass growth in the central region is suppressed; in turn, the central stellar velocity dispersions that are proportional to the mass within the central region do not increase significantly. The difference between $M_{*}-\sigma_{*, z, 10~\rm kpc}$ relations and the VDFs derived from various EAGLE simulations originates from the different star formation history in the central region.

\section{Conclusion} \label{sec:conclusion}

We explored the role of AGN feedback in the stellar kinematics of galaxies based on EAGLE simulations with various AGN feedback models. Our analysis is based on five EAGLE runs (EAGLE-50, EAGLE-eAGN-dT, EAGLE-eAGN-Visc, EAGLE-wAGN, and EAGLE-NoAGN) each covering a cubic volume of (50 Mpc)$^{3}$. We selected pressure-dominated galaxies based on stellar kinematics ($v/\sigma < 0.5$) for log ($M_{*} / \Mdot) > 9$. Furthermore, we constructed the velocity dispersion  complete sample in five EAGLE simulations. We then measured stellar velocity dispersions using the same approach we applied for fiber spectroscopy.

The stellar mass-velocity dispersion  relations derived from the EAGLE simulations show a significant offset compared to observations: simulated galaxies exhibit systematically lower velocity dispersions at fixed stellar mass. This trend is consistent with results from other cosmological simulations, including IllustrisTNG-100 and 300 \citep{Sohn2024a}. As a result, the velocity dispersion functions (VDFs) derived from all five EAGLE runs are significantly lower than the observed VDF.

Despite the systematic offset, comparing the stellar velocity dispersions across EAGLE models with different AGN feedback implementations reveals the impact of AGN feedback on stellar kinematics. The VDF from EAGLE-NoAGN shows a substantial excess of high-velocity dispersion  galaxies ($\log \sigma_{*} > 2.3$) and a flatter slope at $\log \sigma_{*} > 2.0$ relative to those from other EAGLE simulations. This discrepancy strongly suggests that the AGN feedback is certainly responsible for the stellar velocity dispersion (and the underlying mass) distribution of galaxies. 

In order to investigate the physical mechanisms of AGN feedback that affect the stellar kinematics, we investigated the $M_{*} - \sigma_{*}$ scaling relations measured within a fixed 10 kpc aperture in EAGLE simulations. The $M_{*} - \sigma_{*}$ relations in all EAGLE simulations have essentially identical shapes; velocity dispersions are simply proportional to the stellar mass enclosed in the aperture we used for measuring velocity dispersion. In other words, the discrepancy in the VDFs results from the underlying mass distribution rather than any peculiarity in the scaling relation between mass and stellar kinematics. 

We further show that stellar mass concentration varies significantly with AGN feedback. Galaxies in EAGLE-NoAGN display much stronger central mass concentrations (within 10 kpc) than their counterparts in the other runs. This is primarily driven by sustained central star formation in the absence of AGN feedback, as opposed to the suppressed central star formation seen in simulations with AGNs. These differences in stellar mass concentration naturally propagate to differences in stellar velocity dispersions.

The VDF offers a promising probe of the DM subhalo mass function \citep{Sohn2017b, Sohn2024b}. Prior studies \citep{Zahid2018, Sohn2024a} have highlighted a tight correlation between stellar velocity dispersion and the velocity dispersion of the host DM halo. As a next step, comparing the DM mass functions and DM VDFs from simulations will help place further constraints on the stellar VDF as a tracer of the underlying DM distribution.

\begin{acknowledgements}
We thank the referee for carefully reading our manuscript and for the helpful comments that significantly improved it. We acknowledge the Virgo Consortium for making their simulation data available. The EAGLE simulations were performed using the DiRAC-2 facility at Durham, managed by the ICC, and the PRACE facility Curie based in France at TGCC, CEA, Bruy\`{e}res-le-Ch\^{a}tel. J.S. is supported by the National Research Foundation of Korea (NRF) grant funded by the Korean government (MSIT) (RS-2023-00210597). This work was also supported by the Global-LAMP Program of the National Research Foundation of Korea (NRF) grant funded by the Ministry of Education (No. RS-2023-00301976).
\end{acknowledgements}

\bibliographystyle{aa}
\bibliography{aa56640-25}

@ARTICLE{Auger2010,
       author = {{Auger}, M.~W. and {Treu}, T. and {Bolton}, A.~S. and {Gavazzi}, R. and {Koopmans}, L.~V.~E. and {Marshall}, P.~J. and {Moustakas}, L.~A. and {Burles}, S.},
        title = "{The Sloan Lens ACS Survey. X. Stellar, Dynamical, and Total Mass Correlations of Massive Early-type Galaxies}",
      journal = {\apj},
         year = 2010,
        month = nov,
       volume = {724},
       number = {1},
        pages = {511-525},
          doi = {10.1088/0004-637X/724/1/511},
archivePrefix = {arXiv},
       eprint = {1007.2880},
 primaryClass = {astro-ph.CO},
       adsurl = {https://ui.adsabs.harvard.edu/abs/2010ApJ...724..511A}
}

@ARTICLE{Benson2003,
       author = {{Benson}, A.~J. and {Bower}, R.~G. and {Frenk}, C.~S. and {Lacey}, C.~G. and {Baugh}, C.~M. and {Cole}, S.},
        title = "{What Shapes the Luminosity Function of Galaxies?}",
      journal = {\apj},
         year = 2003,
        month = dec,
       volume = {599},
       number = {1},
        pages = {38-49},
          doi = {10.1086/379160},
archivePrefix = {arXiv},
       eprint = {astro-ph/0302450},
 primaryClass = {astro-ph},
       adsurl = {https://ui.adsabs.harvard.edu/abs/2003ApJ...599...38B}
}

@ARTICLE{Bernardi2010,
       author = {{Bernardi}, M. and {Shankar}, F. and {Hyde}, J.~B. and {Mei}, S. and {Marulli}, F. and {Sheth}, R.~K.},
        title = "{Galaxy luminosities, stellar masses, sizes, velocity dispersions as a function of morphological type}",
      journal = {\mnras},
         year = 2010,
        month = jun,
       volume = {404},
       number = {4},
        pages = {2087-2122},
          doi = {10.1111/j.1365-2966.2010.16425.x},
archivePrefix = {arXiv},
       eprint = {0910.1093},
 primaryClass = {astro-ph.CO},
       adsurl = {https://ui.adsabs.harvard.edu/abs/2010MNRAS.404.2087B}
}

@ARTICLE{Bernardi2013,
       author = {{Bernardi}, M. and {Meert}, A. and {Sheth}, R.~K. and {Vikram}, V. and {Huertas-Company}, M. and {Mei}, S. and {Shankar}, F.},
        title = "{The massive end of the luminosity and stellar mass functions: dependence on the fit to the light profile}",
      journal = {\mnras},
         year = 2013,
        month = nov,
       volume = {436},
       number = {1},
        pages = {697-704},
          doi = {10.1093/mnras/stt1607},
archivePrefix = {arXiv},
       eprint = {1304.7778},
 primaryClass = {astro-ph.CO},
       adsurl = {https://ui.adsabs.harvard.edu/abs/2013MNRAS.436..697B}
}

@ARTICLE{Bernardi2017,
       author = {{Bernardi}, M. and {Meert}, A. and {Sheth}, R.~K. and {Fischer}, J.-L. and {Huertas-Company}, M. and {Maraston}, C. and {Shankar}, F. and {Vikram}, V.},
        title = "{The high mass end of the stellar mass function: Dependence on stellar population models and agreement between fits to the light profile}",
      journal = {\mnras},
         year = 2017,
        month = may,
       volume = {467},
       number = {2},
        pages = {2217-2233},
          doi = {10.1093/mnras/stx176},
archivePrefix = {arXiv},
       eprint = {1604.01036},
 primaryClass = {astro-ph.GA},
       adsurl = {https://ui.adsabs.harvard.edu/abs/2017MNRAS.467.2217B}
}

@ARTICLE{Binney2004,
       author = {{Binney}, James},
        title = "{On the origin of the galaxy luminosity function}",
      journal = {\mnras},
         year = 2004,
        month = feb,
       volume = {347},
       number = {4},
        pages = {1093-1096},
          doi = {10.1111/j.1365-2966.2004.07277.x},
archivePrefix = {arXiv},
       eprint = {astro-ph/0308172},
 primaryClass = {astro-ph},
       adsurl = {https://ui.adsabs.harvard.edu/abs/2004MNRAS.347.1093B}
}

@ARTICLE{Bogdan2015,
       author = {{Bogd{\'a}n}, {\'A}kos and {Goulding}, Andy D.},
        title = "{Connecting Dark Matter Halos with the Galaxy Center and the Supermassive Black Hole}",
      journal = {\apj},
         year = 2015,
        month = feb,
       volume = {800},
       number = {2},
          eid = {124},
        pages = {124},
          doi = {10.1088/0004-637X/800/2/124},
archivePrefix = {arXiv},
       eprint = {1502.05043},
 primaryClass = {astro-ph.GA},
       adsurl = {https://ui.adsabs.harvard.edu/abs/2015ApJ...800..124B}
}

@ARTICLE{Cannarozzo2020,
       author = {{Cannarozzo}, Carlo and {Sonnenfeld}, Alessandro and {Nipoti}, Carlo},
        title = "{The cosmic evolution of the stellar mass-velocity dispersion relation of early-type galaxies}",
      journal = {\mnras},
         year = 2020,
        month = oct,
       volume = {498},
       number = {1},
        pages = {1101-1120},
          doi = {10.1093/mnras/staa2147},
archivePrefix = {arXiv},
       eprint = {1910.06987},
 primaryClass = {astro-ph.GA},
       adsurl = {https://ui.adsabs.harvard.edu/abs/2020MNRAS.498.1101C}
}

@ARTICLE{Cappellari2013a,
       author = {{Cappellari}, Michele and {Scott}, Nicholas and {Alatalo}, Katherine and {Blitz}, Leo and {Bois}, Maxime and {Bournaud}, Fr{\'e}d{\'e}ric and {Bureau}, M. and {Crocker}, Alison F. and {Davies}, Roger L. and {Davis}, Timothy A. and {de Zeeuw}, P.~T. and {Duc}, Pierre-Alain and {Emsellem}, Eric and {Khochfar}, Sadegh and {Krajnovi{\'c}}, Davor and {Kuntschner}, Harald and {McDermid}, Richard M. and {Morganti}, Raffaella and {Naab}, Thorsten and {Oosterloo}, Tom and {Sarzi}, Marc and {Serra}, Paolo and {Weijmans}, Anne-Marie and {Young}, Lisa M.},
        title = "{The ATLAS$^{3D}$ project - XV. Benchmark for early-type galaxies scaling relations from 260 dynamical models: mass-to-light ratio, dark matter, Fundamental Plane and Mass Plane}",
      journal = {\mnras},
         year = 2013,
        month = jul,
       volume = {432},
       number = {3},
        pages = {1709-1741},
          doi = {10.1093/mnras/stt562},
archivePrefix = {arXiv},
       eprint = {1208.3522},
 primaryClass = {astro-ph.CO},
       adsurl = {https://ui.adsabs.harvard.edu/abs/2013MNRAS.432.1709C}
}

@ARTICLE{Cappellari2016,
       author = {{Cappellari}, Michele},
        title = "{Structure and Kinematics of Early-Type Galaxies from Integral Field Spectroscopy}",
      journal = {\araa},
         year = 2016,
        month = sep,
       volume = {54},
        pages = {597-665},
          doi = {10.1146/annurev-astro-082214-122432},
archivePrefix = {arXiv},
       eprint = {1602.04267},
 primaryClass = {astro-ph.GA},
       adsurl = {https://ui.adsabs.harvard.edu/abs/2016ARA&A..54..597C}
}

@ARTICLE{Chae2010,
       author = {{Chae}, Kyu-Hyun},
        title = "{Galaxy evolution from strong-lensing statistics: the differential evolution of the velocity dispersion function in concord with the {\ensuremath{\Lambda}} cold dark matter paradigm}",
      journal = {\mnras},
         year = 2010,
        month = mar,
       volume = {402},
       number = {3},
        pages = {2031-2048},
          doi = {10.1111/j.1365-2966.2009.16073.x},
archivePrefix = {arXiv},
       eprint = {0811.0560},
 primaryClass = {astro-ph},
       adsurl = {https://ui.adsabs.harvard.edu/abs/2010MNRAS.402.2031C}
}

@ARTICLE{Choi2018,
       author = {{Choi}, Ena and {Somerville}, Rachel S. and {Ostriker}, Jeremiah P. and {Naab}, Thorsten and {Hirschmann}, Michaela},
        title = "{The Role of Black Hole Feedback on Size and Structural Evolution in Massive Galaxies}",
      journal = {\apj},
         year = 2018,
        month = oct,
       volume = {866},
       number = {2},
          eid = {91},
        pages = {91},
          doi = {10.3847/1538-4357/aae076},
archivePrefix = {arXiv},
       eprint = {1809.02143},
 primaryClass = {astro-ph.GA},
       adsurl = {https://ui.adsabs.harvard.edu/abs/2018ApJ...866...91C}
}

@ARTICLE{Choi2007,
       author = {{Choi}, Yun-Young and {Park}, Changbom and {Vogeley}, Michael S.},
        title = "{Internal and Collective Properties of Galaxies in the Sloan Digital Sky Survey}",
      journal = {\apj},
         year = 2007,
        month = apr,
       volume = {658},
       number = {2},
        pages = {884-897},
          doi = {10.1086/511060},
archivePrefix = {arXiv},
       eprint = {astro-ph/0611607},
 primaryClass = {astro-ph},
       adsurl = {https://ui.adsabs.harvard.edu/abs/2007ApJ...658..884C}
}

@ARTICLE{Ciotti1997,
       author = {{Ciotti}, Luca and {Ostriker}, Jeremiah P.},
        title = "{Cooling Flows and Quasars: Different Aspects of the Same Phenomenon? I. Concepts}",
      journal = {\apjl},
         year = 1997,
        month = oct,
       volume = {487},
       number = {2},
        pages = {L105-L108},
          doi = {10.1086/310902},
archivePrefix = {arXiv},
       eprint = {astro-ph/9706281},
 primaryClass = {astro-ph},
       adsurl = {https://ui.adsabs.harvard.edu/abs/1997ApJ...487L.105C}
}

@INPROCEEDINGS{Combes2017,
       author = {{Combes}, Francoise},
        title = "{AGN feedback and its quenching efficiency}",
    booktitle = {Quasars at all Cosmic Epochs},
         year = 2017,
        month = apr,
          eid = {20},
        pages = {20},
          doi = {10.5281/zenodo.569791},
       adsurl = {https://ui.adsabs.harvard.edu/abs/2017qace.confE..20C}
}

@ARTICLE{Conroy2009,
       author = {{Conroy}, Charlie and {Gunn}, James E. and {White}, Martin},
        title = "{The Propagation of Uncertainties in Stellar Population Synthesis Modeling. I. The Relevance of Uncertain Aspects of Stellar Evolution and the Initial Mass Function to the Derived Physical Properties of Galaxies}",
      journal = {\apj},
         year = 2009,
        month = jul,
       volume = {699},
       number = {1},
        pages = {486-506},
          doi = {10.1088/0004-637X/699/1/486},
archivePrefix = {arXiv},
       eprint = {0809.4261},
 primaryClass = {astro-ph},
       adsurl = {https://ui.adsabs.harvard.edu/abs/2009ApJ...699..486C}
}

@ARTICLE{Crain2015,
       author = {{Crain}, Robert A. and {Schaye}, Joop and {Bower}, Richard G. and {Furlong}, Michelle and {Schaller}, Matthieu and {Theuns}, Tom and {Dalla Vecchia}, Claudio and {Frenk}, Carlos S. and {McCarthy}, Ian G. and {Helly}, John C. and {Jenkins}, Adrian and {Rosas-Guevara}, Yetli M. and {White}, Simon D.~M. and {Trayford}, James W.},
        title = "{The EAGLE simulations of galaxy formation: calibration of subgrid physics and model variations}",
      journal = {\mnras},
         year = 2015,
        month = jun,
       volume = {450},
       number = {2},
        pages = {1937-1961},
          doi = {10.1093/mnras/stv725},
archivePrefix = {arXiv},
       eprint = {1501.01311},
 primaryClass = {astro-ph.GA},
       adsurl = {https://ui.adsabs.harvard.edu/abs/2015MNRAS.450.1937C}
}

@ARTICLE{Djorgovski1987,
       author = {{Djorgovski}, S. and {Davis}, Marc},
        title = "{Fundamental Properties of Elliptical Galaxies}",
      journal = {\apj},
         year = 1987,
        month = feb,
       volume = {313},
        pages = {59},
          doi = {10.1086/164948},
       adsurl = {https://ui.adsabs.harvard.edu/abs/1987ApJ...313...59D}
}

@ARTICLE{Dolag2009,
       author = {{Dolag}, K. and {Borgani}, S. and {Murante}, G. and {Springel}, V.},
        title = "{Substructures in hydrodynamical cluster simulations}",
      journal = {\mnras},
         year = 2009,
        month = oct,
       volume = {399},
       number = {2},
        pages = {497-514},
          doi = {10.1111/j.1365-2966.2009.15034.x},
archivePrefix = {arXiv},
       eprint = {0808.3401},
 primaryClass = {astro-ph},
       adsurl = {https://ui.adsabs.harvard.edu/abs/2009MNRAS.399..497D}
}

@ARTICLE{Dubois2021,
       author = {{Dubois}, Yohan and {Beckmann}, Ricarda and {Bournaud}, Fr{\'e}d{\'e}ric and {Choi}, Hoseung and {Devriendt}, Julien and {Jackson}, Ryan and {Kaviraj}, Sugata and {Kimm}, Taysun and {Kraljic}, Katarina and {Laigle}, Clotilde and {Martin}, Garreth and {Park}, Min-Jung and {Peirani}, S{\'e}bastien and {Pichon}, Christophe and {Volonteri}, Marta and {Yi}, Sukyoung K.},
        title = "{Introducing the NEWHORIZON simulation: Galaxy properties with resolved internal dynamics across cosmic time}",
      journal = {\aap},
         year = 2021,
        month = jul,
       volume = {651},
          eid = {A109},
        pages = {A109},
          doi = {10.1051/0004-6361/202039429},
archivePrefix = {arXiv},
       eprint = {2009.10578},
 primaryClass = {astro-ph.GA},
       adsurl = {https://ui.adsabs.harvard.edu/abs/2021A&A...651A.109D}
}

@ARTICLE{Dubois2016,
       author = {{Dubois}, Yohan and {Peirani}, S{\'e}bastien and {Pichon}, Christophe and {Devriendt}, Julien and {Gavazzi}, Rapha{\"e}l and {Welker}, Charlotte and {Volonteri}, Marta},
        title = "{The HORIZON-AGN simulation: morphological diversity of galaxies promoted by AGN feedback}",
      journal = {\mnras},
         year = 2016,
        month = dec,
       volume = {463},
       number = {4},
        pages = {3948-3964},
          doi = {10.1093/mnras/stw2265},
archivePrefix = {arXiv},
       eprint = {1606.03086},
 primaryClass = {astro-ph.GA},
       adsurl = {https://ui.adsabs.harvard.edu/abs/2016MNRAS.463.3948D}
}

@ARTICLE{Faber1976,
       author = {{Faber}, S.~M. and {Jackson}, R.~E.},
        title = "{Velocity dispersions and mass-to-light ratios for elliptical galaxies.}",
      journal = {\apj},
         year = 1976,
        month = mar,
       volume = {204},
        pages = {668-683},
          doi = {10.1086/154215},
       adsurl = {https://ui.adsabs.harvard.edu/abs/1976ApJ...204..668F}
}

@ARTICLE{Fan2008,
       author = {{Fan}, L. and {Lapi}, A. and {De Zotti}, G. and {Danese}, L.},
        title = "{The Dramatic Size Evolution of Elliptical Galaxies and the Quasar Feedback}",
      journal = {\apjl},
         year = 2008,
        month = dec,
       volume = {689},
       number = {2},
        pages = {L101},
          doi = {10.1086/595784},
archivePrefix = {arXiv},
       eprint = {0809.4574},
 primaryClass = {astro-ph},
       adsurl = {https://ui.adsabs.harvard.edu/abs/2008ApJ...689L.101F}
}

@ARTICLE{Fan2010,
       author = {{Fan}, L. and {Lapi}, A. and {Bressan}, A. and {Bernardi}, M. and {De Zotti}, G. and {Danese}, L.},
        title = "{Cosmic Evolution of Size and Velocity Dispersion for Early-type Galaxies}",
      journal = {\apj},
         year = 2010,
        month = aug,
       volume = {718},
       number = {2},
        pages = {1460-1475},
          doi = {10.1088/0004-637X/718/2/1460},
archivePrefix = {arXiv},
       eprint = {1006.2303},
 primaryClass = {astro-ph.CO},
       adsurl = {https://ui.adsabs.harvard.edu/abs/2010ApJ...718.1460F}
}

@ARTICLE{Ferragamo2021,
       author = {{Ferragamo}, A. and {Barrena}, R. and {Rubi{\~n}o-Mart{\'\i}n}, J.~A. and {Aguado-Barahona}, A. and {Streblyanska}, A. and {Tramonte}, D. and {G{\'e}nova-Santos}, R.~T. and {Hempel}, A. and {Lietzen}, H.},
        title = "{Velocity dispersion and dynamical mass for 270 galaxy clusters in the Planck PSZ1 catalogue}",
      journal = {\aap},
         year = 2021,
        month = nov,
       volume = {655},
          eid = {A115},
        pages = {A115},
          doi = {10.1051/0004-6361/202140382},
archivePrefix = {arXiv},
       eprint = {2109.04967},
 primaryClass = {astro-ph.CO},
       adsurl = {https://ui.adsabs.harvard.edu/abs/2021A&A...655A.115F}
}

@ARTICLE{Furlong2017,
       author = {{Furlong}, M. and {Bower}, R.~G. and {Crain}, R.~A. and {Schaye}, J. and {Theuns}, T. and {Trayford}, J.~W. and {Qu}, Y. and {Schaller}, M. and {Berthet}, M. and {Helly}, J.~C.},
        title = "{Size evolution of normal and compact galaxies in the EAGLE simulation}",
      journal = {\mnras},
         year = 2017,
        month = feb,
       volume = {465},
       number = {1},
        pages = {722-738},
          doi = {10.1093/mnras/stw2740},
archivePrefix = {arXiv},
       eprint = {1510.05645},
 primaryClass = {astro-ph.GA},
       adsurl = {https://ui.adsabs.harvard.edu/abs/2017MNRAS.465..722F}
}

@ARTICLE{Goubert2024,
       author = {{Goubert}, Paul H. and {Bluck}, Asa F.~L. and {Piotrowska}, Joanna M. and {Maiolino}, Roberto},
        title = "{The role of environment and AGN feedback in quenching local galaxies: comparing cosmological hydrodynamical simulations to the SDSS}",
      journal = {\mnras},
         year = 2024,
        month = mar,
       volume = {528},
       number = {3},
        pages = {4891-4921},
          doi = {10.1093/mnras/stae269},
archivePrefix = {arXiv},
       eprint = {2401.12953},
 primaryClass = {astro-ph.GA},
       adsurl = {https://ui.adsabs.harvard.edu/abs/2024MNRAS.528.4891G}
}

@ARTICLE{Grillo2008,
       author = {{Grillo}, C. and {Lombardi}, M. and {Bertin}, G.},
        title = "{Cosmological parameters from strong gravitational lensing and stellar dynamics in elliptical galaxies}",
      journal = {\aap},
         year = 2008,
        month = jan,
       volume = {477},
       number = {2},
        pages = {397-406},
          doi = {10.1051/0004-6361:20077534},
archivePrefix = {arXiv},
       eprint = {0711.0882},
 primaryClass = {astro-ph},
       adsurl = {https://ui.adsabs.harvard.edu/abs/2008A&A...477..397G}
}

@ARTICLE{Harrison2024,
       author = {{Harrison}, Chris M. and {Ramos Almeida}, Cristina},
        title = "{Observational Tests of Active Galactic Nuclei Feedback: An Overview of Approaches and Interpretation}",
      journal = {Galaxies},
         year = 2024,
        month = apr,
       volume = {12},
       number = {2},
          eid = {17},
        pages = {17},
          doi = {10.3390/galaxies12020017},
archivePrefix = {arXiv},
       eprint = {2404.08050},
 primaryClass = {astro-ph.GA},
       adsurl = {https://ui.adsabs.harvard.edu/abs/2024Galax..12...17H}
}

@ARTICLE{Hopkins2010,
       author = {{Hopkins}, Philip F. and {Quataert}, Eliot},
        title = "{How do massive black holes get their gas?}",
      journal = {\mnras},
         year = 2010,
        month = sep,
       volume = {407},
       number = {3},
        pages = {1529-1564},
          doi = {10.1111/j.1365-2966.2010.17064.x},
archivePrefix = {arXiv},
       eprint = {0912.3257},
 primaryClass = {astro-ph.CO},
       adsurl = {https://ui.adsabs.harvard.edu/abs/2010MNRAS.407.1529H}
}

@ARTICLE{Knabel2025,
       author = {{Knabel}, Shawn and {Mozumdar}, Pritom and {Shajib}, Anowar J. and {Treu}, Tommaso and {Cappellari}, Michele and {Spiniello}, Chiara and {Birrer}, Simon},
        title = "{TDCOSMO: XIX. Measuring stellar velocity dispersion with sub-percent accuracy for cosmography}",
      journal = {\aap},
         year = 2025,
        month = nov,
       volume = {703},
          eid = {A117},
        pages = {A117},
          doi = {10.1051/0004-6361/202554229},
archivePrefix = {arXiv},
       eprint = {2502.16034},
 primaryClass = {astro-ph.GA},
       adsurl = {https://ui.adsabs.harvard.edu/abs/2025A&A...703A.117K}
}

@ARTICLE{Li2009,
       author = {{Li}, Cheng and {White}, Simon D.~M.},
        title = "{The distribution of stellar mass in the low-redshift Universe}",
      journal = {\mnras},
         year = 2009,
        month = oct,
       volume = {398},
       number = {4},
        pages = {2177-2187},
          doi = {10.1111/j.1365-2966.2009.15268.x},
archivePrefix = {arXiv},
       eprint = {0901.0706},
 primaryClass = {astro-ph.CO},
       adsurl = {https://ui.adsabs.harvard.edu/abs/2009MNRAS.398.2177L}
}

@ARTICLE{Li2018,
       author = {{Li}, Ya-Ping and {Yuan}, Feng and {Mo}, Houjun and {Yoon}, Doosoo and {Gan}, Zhaoming and {Ho}, Luis C. and {Wang}, Bo and {Ostriker}, Jeremiah P. and {Ciotti}, Luca},
        title = "{Stellar and AGN Feedback in Isolated Early-type Galaxies: The Role in Regulating Star Formation and ISM Properties}",
      journal = {\apj},
         year = 2018,
        month = oct,
       volume = {866},
       number = {1},
          eid = {70},
        pages = {70},
          doi = {10.3847/1538-4357/aade8b},
archivePrefix = {arXiv},
       eprint = {1803.01444},
 primaryClass = {astro-ph.HE},
       adsurl = {https://ui.adsabs.harvard.edu/abs/2018ApJ...866...70L}
}

@ARTICLE{McAlpine2016,
       author = {{McAlpine}, S. and {Helly}, J.~C. and {Schaller}, M. and {Trayford}, J.~W. and {Qu}, Y. and {Furlong}, M. and {Bower}, R.~G. and {Crain}, R.~A. and {Schaye}, J. and {Theuns}, T. and {Dalla Vecchia}, C. and {Frenk}, C.~S. and {McCarthy}, I.~G. and {Jenkins}, A. and {Rosas-Guevara}, Y. and {White}, S.~D.~M. and {Baes}, M. and {Camps}, P. and {Lemson}, G.},
        title = "{The EAGLE simulations of galaxy formation: Public release of halo and galaxy catalogues}",
      journal = {Astronomy and Computing},
         year = 2016,
        month = apr,
       volume = {15},
        pages = {72-89},
          doi = {10.1016/j.ascom.2016.02.004},
archivePrefix = {arXiv},
       eprint = {1510.01320},
 primaryClass = {astro-ph.GA},
       adsurl = {https://ui.adsabs.harvard.edu/abs/2016A&C....15...72M}
}

@ARTICLE{Menci2002,
       author = {{Menci}, N. and {Cavaliere}, A. and {Fontana}, A. and {Giallongo}, E. and {Poli}, F.},
        title = "{Binary Aggregations in Hierarchical Galaxy Formation: The Evolution of the Galaxy Luminosity Function}",
      journal = {\apj},
         year = 2002,
        month = aug,
       volume = {575},
       number = {1},
        pages = {18-32},
          doi = {10.1086/341191},
archivePrefix = {arXiv},
       eprint = {astro-ph/0204178},
 primaryClass = {astro-ph},
       adsurl = {https://ui.adsabs.harvard.edu/abs/2002ApJ...575...18M}
}

@ARTICLE{Mogotsi2019,
       author = {{Mogotsi}, Keoikantse Moses and {Romeo}, Alessandro B.},
        title = "{The stellar velocity dispersion in nearby spirals: radial profiles and correlations}",
      journal = {\mnras},
         year = 2019,
        month = nov,
       volume = {489},
       number = {3},
        pages = {3797-3809},
          doi = {10.1093/mnras/stz2370},
archivePrefix = {arXiv},
       eprint = {1804.10119},
 primaryClass = {astro-ph.GA},
       adsurl = {https://ui.adsabs.harvard.edu/abs/2019MNRAS.489.3797M}
}

@ARTICLE{Montero-Dorta2017,
       author = {{Montero-Dorta}, Antonio D. and {Bolton}, Adam S. and {Shu}, Yiping},
        title = "{A direct measurement of the high-mass end of the velocity dispersion function at z {\ensuremath{\sim}} 0.55 from SDSS-III/BOSS}",
      journal = {\mnras},
         year = 2017,
        month = jun,
       volume = {468},
       number = {1},
        pages = {47-58},
          doi = {10.1093/mnras/stx321},
archivePrefix = {arXiv},
       eprint = {1607.06820},
 primaryClass = {astro-ph.GA},
       adsurl = {https://ui.adsabs.harvard.edu/abs/2017MNRAS.468...47M}
}

@ARTICLE{Nigoche-Netro2011,
       author = {{Nigoche-Netro}, A. and {Aguerri}, J.~A.~L. and {Lagos}, P. and {Ruelas-Mayorga}, A. and {S{\'a}nchez}, L.~J. and {Mu{\~n}oz-Tu{\~n}{\'o}n}, C. and {Machado}, A.},
        title = "{The intrinsic dispersion in the Faber-Jackson relation for early-type galaxies as function of the mass and redshift}",
      journal = {\aap},
         year = 2011,
        month = oct,
       volume = {534},
          eid = {A61},
        pages = {A61},
          doi = {10.1051/0004-6361/201016360},
archivePrefix = {arXiv},
       eprint = {1107.6017},
 primaryClass = {astro-ph.CO},
       adsurl = {https://ui.adsabs.harvard.edu/abs/2011A&A...534A..61N}
}

@ARTICLE{Pillepich2018,
       author = {{Pillepich}, Annalisa and {Springel}, Volker and {Nelson}, Dylan and {Genel}, Shy and {Naiman}, Jill and {Pakmor}, R{\"u}diger and {Hernquist}, Lars and {Torrey}, Paul and {Vogelsberger}, Mark and {Weinberger}, Rainer and {Marinacci}, Federico},
        title = "{Simulating galaxy formation with the IllustrisTNG model}",
      journal = {\mnras},
         year = 2018,
        month = jan,
       volume = {473},
       number = {3},
        pages = {4077-4106},
          doi = {10.1093/mnras/stx2656},
archivePrefix = {arXiv},
       eprint = {1703.02970},
 primaryClass = {astro-ph.GA},
       adsurl = {https://ui.adsabs.harvard.edu/abs/2018MNRAS.473.4077P}
}

@ARTICLE{PlanckCollaboration2014,
       author = {{Planck Collaboration} and {Ade}, P.~A.~R. and {Aghanim}, N. and {Alves}, M.~I.~R. and {Armitage-Caplan}, C. and {Arnaud}, M. and {Ashdown}, M. and {Atrio-Barandela}, F. and {Aumont}, J. and {Aussel}, H. and {Baccigalupi}, C. and {Banday}, A.~J. and {Barreiro}, R.~B. and {Barrena}, R. and {Bartelmann}, M. and {Bartlett}, J.~G. and {Bartolo}, N. and {Basak}, S. and {Battaner}, E. and {Battye}, R. and {Benabed}, K. and {Beno{\^\i}t}, A. and {Benoit-L{\'e}vy}, A. and {Bernard}, J.-P. and {Bersanelli}, M. and {Bertincourt}, B. and {Bethermin}, M. and {Bielewicz}, P. and {Bikmaev}, I. and {Blanchard}, A. and {Bobin}, J. and {Bock}, J.~J. and {B{\"o}hringer}, H. and {Bonaldi}, A. and {Bonavera}, L. and {Bond}, J.~R. and {Borrill}, J. and {Bouchet}, F.~R. and {Boulanger}, F. and {Bourdin}, H. and {Bowyer}, J.~W. and {Bridges}, M. and {Brown}, M.~L. and {Bucher}, M. and {Burenin}, R. and {Burigana}, C. and {Butler}, R.~C. and {Calabrese}, E. and {Cappellini}, B. and {Cardoso}, J.-F. and {Carr}, R. and {Carvalho}, P. and {Casale}, M. and {Castex}, G. and {Catalano}, A. and {Challinor}, A. and {Chamballu}, A. and {Chary}, R.-R. and {Chen}, X. and {Chiang}, H.~C. and {Chiang}, L.-Y. and {Chon}, G. and {Christensen}, P.~R. and {Churazov}, E. and {Church}, S. and {Clemens}, M. and {Clements}, D.~L. and {Colombi}, S. and {Colombo}, L.~P.~L. and {Combet}, C. and {Comis}, B. and {Couchot}, F. and {Coulais}, A. and {Crill}, B.~P. and {Cruz}, M. and {Curto}, A. and {Cuttaia}, F. and {Da Silva}, A. and {Dahle}, H. and {Danese}, L. and {Davies}, R.~D. and {Davis}, R.~J. and {de Bernardis}, P. and {de Rosa}, A. and {de Zotti}, G. and {D{\'e}chelette}, T. and {Delabrouille}, J. and {Delouis}, J.-M. and {D{\'e}mocl{\`e}s}, J. and {D{\'e}sert}, F.-X. and {Dick}, J. and {Dickinson}, C. and {Diego}, J.~M. and {Dolag}, K. and {Dole}, H. and {Donzelli}, S. and {Dor{\'e}}, O. and {Douspis}, M. and {Ducout}, A. and {Dunkley}, J. and {Dupac}, X. and {Efstathiou}, G. and {Elsner}, F. and {En{\ss}lin}, T.~A. and {Eriksen}, H.~K. and {Fabre}, O. and {Falgarone}, E. and {Falvella}, M.~C. and {Fantaye}, Y. and {Fergusson}, J. and {Filliard}, C. and {Finelli}, F. and {Flores-Cacho}, I. and {Foley}, S. and {Forni}, O. and {Fosalba}, P. and {Frailis}, M. and {Fraisse}, A.~A. and {Franceschi}, E. and {Freschi}, M. and {Fromenteau}, S. and {Frommert}, M. and {Gaier}, T.~C. and {Galeotta}, S. and {Gallegos}, J. and {Galli}, S. and {Gandolfo}, B. and {Ganga}, K. and {Gauthier}, C. and {G{\'e}nova-Santos}, R.~T. and {Ghosh}, T. and {Giard}, M. and {Giardino}, G. and {Gilfanov}, M. and {Girard}, D. and {Giraud-H{\'e}raud}, Y. and {Gjerl{\o}w}, E. and {Gonz{\'a}lez-Nuevo}, J. and {G{\'o}rski}, K.~M. and {Gratton}, S. and {Gregorio}, A. and {Gruppuso}, A. and {Gudmundsson}, J.~E. and {Haissinski}, J. and {Hamann}, J. and {Hansen}, F.~K. and {Hansen}, M. and {Hanson}, D. and {Harrison}, D.~L. and {Heavens}, A. and {Helou}, G. and {Hempel}, A. and {Henrot-Versill{\'e}}, S. and {Hern{\'a}ndez-Monteagudo}, C. and {Herranz}, D. and {Hildebrandt}, S.~R. and {Hivon}, E. and {Ho}, S. and {Hobson}, M. and {Holmes}, W.~A. and {Hornstrup}, A. and {Hou}, Z. and {Hovest}, W. and {Huey}, G. and {Huffenberger}, K.~M. and {Hurier}, G. and {Ili{\'c}}, S. and {Jaffe}, A.~H. and {Jaffe}, T.~R. and {Jasche}, J. and {Jewell}, J. and {Jones}, W.~C. and {Juvela}, M. and {Kalberla}, P. and {Kangaslahti}, P. and {Keih{\"a}nen}, E. and {Kerp}, J. and {Keskitalo}, R. and {Khamitov}, I. and {Kiiveri}, K. and {Kim}, J. and {Kisner}, T.~S. and {Kneissl}, R. and {Knoche}, J. and {Knox}, L. and {Kunz}, M. and {Kurki-Suonio}, H. and {Lacasa}, F. and {Lagache}, G. and {L{\"a}hteenm{\"a}ki}, A. and {Lamarre}, J.-M. and {Langer}, M. and {Lasenby}, A. and {Lattanzi}, M. and {Laureijs}, R.~J. and {Lavabre}, A. and {Lawrence}, C.~R. and {Le Jeune}, M. and {Leach}, S. and {Leahy}, J.~P.},
        title = "{Planck 2013 results. I. Overview of products and scientific results}",
      journal = {\aap},
         year = 2014,
        month = nov,
       volume = {571},
          eid = {A1},
        pages = {A1},
          doi = {10.1051/0004-6361/201321529},
archivePrefix = {arXiv},
       eprint = {1303.5062},
 primaryClass = {astro-ph.CO},
       adsurl = {https://ui.adsabs.harvard.edu/abs/2014A&A...571A...1P}
}

@ARTICLE{Rosas-Guevara2015,
       author = {{Rosas-Guevara}, Y.~M. and {Bower}, R.~G. and {Schaye}, J. and {Furlong}, M. and {Frenk}, C.~S. and {Booth}, C.~M. and {Crain}, R.~A. and {Dalla Vecchia}, C. and {Schaller}, M. and {Theuns}, T.},
        title = "{The impact of angular momentum on black hole accretion rates in simulations of galaxy formation}",
      journal = {\mnras},
         year = 2015,
        month = nov,
       volume = {454},
       number = {1},
        pages = {1038-1057},
          doi = {10.1093/mnras/stv2056},
archivePrefix = {arXiv},
       eprint = {1312.0598},
 primaryClass = {astro-ph.CO},
       adsurl = {https://ui.adsabs.harvard.edu/abs/2015MNRAS.454.1038R}
}

@ARTICLE{Sales2010,
       author = {{Sales}, Laura V. and {Navarro}, Julio F. and {Schaye}, Joop and {Dalla Vecchia}, Claudio and {Springel}, Volker and {Booth}, C.~M.},
        title = "{Feedback and the structure of simulated galaxies at redshift z= 2}",
      journal = {\mnras},
         year = 2010,
        month = dec,
       volume = {409},
       number = {4},
        pages = {1541-1556},
          doi = {10.1111/j.1365-2966.2010.17391.x},
archivePrefix = {arXiv},
       eprint = {1004.5386},
 primaryClass = {astro-ph.CO},
       adsurl = {https://ui.adsabs.harvard.edu/abs/2010MNRAS.409.1541S}
}

@ARTICLE{Sales2012,
       author = {{Sales}, Laura V. and {Navarro}, Julio F. and {Theuns}, Tom and {Schaye}, Joop and {White}, Simon D.~M. and {Frenk}, Carlos S. and {Crain}, Robert A. and {Dalla Vecchia}, Claudio},
        title = "{The origin of discs and spheroids in simulated galaxies}",
      journal = {\mnras},
         year = 2012,
        month = jun,
       volume = {423},
       number = {2},
        pages = {1544-1555},
          doi = {10.1111/j.1365-2966.2012.20975.x},
archivePrefix = {arXiv},
       eprint = {1112.2220},
 primaryClass = {astro-ph.CO},
       adsurl = {https://ui.adsabs.harvard.edu/abs/2012MNRAS.423.1544S}
}

@ARTICLE{Schawinski2007,
       author = {{Schawinski}, Kevin and {Thomas}, Daniel and {Sarzi}, Marc and {Maraston}, Claudia and {Kaviraj}, Sugata and {Joo}, Seok-Joo and {Yi}, Sukyoung K. and {Silk}, Joseph},
        title = "{Observational evidence for AGN feedback in early-type galaxies}",
      journal = {\mnras},
         year = 2007,
        month = dec,
       volume = {382},
       number = {4},
        pages = {1415-1431},
          doi = {10.1111/j.1365-2966.2007.12487.x},
archivePrefix = {arXiv},
       eprint = {0709.3015},
 primaryClass = {astro-ph},
       adsurl = {https://ui.adsabs.harvard.edu/abs/2007MNRAS.382.1415S}
}

@ARTICLE{Schaye2015,
       author = {{Schaye}, Joop and {Crain}, Robert A. and {Bower}, Richard G. and {Furlong}, Michelle and {Schaller}, Matthieu and {Theuns}, Tom and {Dalla Vecchia}, Claudio and {Frenk}, Carlos S. and {McCarthy}, I.~G. and {Helly}, John C. and {Jenkins}, Adrian and {Rosas-Guevara}, Y.~M. and {White}, Simon D.~M. and {Baes}, Maarten and {Booth}, C.~M. and {Camps}, Peter and {Navarro}, Julio F. and {Qu}, Yan and {Rahmati}, Alireza and {Sawala}, Till and {Thomas}, Peter A. and {Trayford}, James},
        title = "{The EAGLE project: simulating the evolution and assembly of galaxies and their environments}",
      journal = {\mnras},
         year = 2015,
        month = jan,
       volume = {446},
       number = {1},
        pages = {521-554},
          doi = {10.1093/mnras/stu2058},
archivePrefix = {arXiv},
       eprint = {1407.7040},
 primaryClass = {astro-ph.GA},
       adsurl = {https://ui.adsabs.harvard.edu/abs/2015MNRAS.446..521S}
}

@ARTICLE{Sheth2003,
       author = {{Sheth}, Ravi K. and {Bernardi}, Mariangela and {Schechter}, Paul L. and {Burles}, Scott and {Eisenstein}, Daniel J. and {Finkbeiner}, Douglas P. and {Frieman}, Joshua and {Lupton}, Robert H. and {Schlegel}, David J. and {Subbarao}, Mark and {Shimasaku}, K. and {Bahcall}, Neta A. and {Brinkmann}, J. and {Ivezi{\'c}}, {\v{Z}}eljko},
        title = "{The Velocity Dispersion Function of Early-Type Galaxies}",
      journal = {\apj},
         year = 2003,
        month = sep,
       volume = {594},
       number = {1},
        pages = {225-231},
          doi = {10.1086/376794},
archivePrefix = {arXiv},
       eprint = {astro-ph/0303092},
 primaryClass = {astro-ph},
       adsurl = {https://ui.adsabs.harvard.edu/abs/2003ApJ...594..225S}
}

@ARTICLE{Silk2005,
       author = {{Silk}, Joseph},
        title = "{Ultraluminous starbursts from supermassive black hole-induced outflows}",
      journal = {\mnras},
         year = 2005,
        month = dec,
       volume = {364},
       number = {4},
        pages = {1337-1342},
          doi = {10.1111/j.1365-2966.2005.09672.x},
archivePrefix = {arXiv},
       eprint = {astro-ph/0509149},
 primaryClass = {astro-ph},
       adsurl = {https://ui.adsabs.harvard.edu/abs/2005MNRAS.364.1337S}
}

@ARTICLE{Silk1998,
       author = {{Silk}, Joseph and {Rees}, Martin J.},
        title = "{Quasars and galaxy formation}",
      journal = {\aap},
         year = 1998,
        month = mar,
       volume = {331},
        pages = {L1-L4},
          doi = {10.48550/arXiv.astro-ph/9801013},
archivePrefix = {arXiv},
       eprint = {astro-ph/9801013},
 primaryClass = {astro-ph},
       adsurl = {https://ui.adsabs.harvard.edu/abs/1998A&A...331L...1S}
}

@ARTICLE{Sohn2024a,
       author = {{Sohn}, Jubee and {Geller}, Margaret J. and {Borrow}, Josh and {Vogelsberger}, Mark},
        title = "{Velocity Dispersions of Quiescent Galaxies in IllustrisTNG}",
      journal = {\apj},
         year = 2024,
        month = apr,
       volume = {964},
       number = {2},
          eid = {178},
        pages = {178},
          doi = {10.3847/1538-4357/ad2c0a},
archivePrefix = {arXiv},
       eprint = {2402.14218},
 primaryClass = {astro-ph.GA},
       adsurl = {https://ui.adsabs.harvard.edu/abs/2024ApJ...964..178S}
}

@ARTICLE{Sohn2017a,
       author = {{Sohn}, Jubee and {Geller}, Margaret J. and {Zahid}, H. Jabran and {Fabricant}, Daniel G. and {Diaferio}, Antonaldo and {Rines}, Kenneth J.},
        title = "{The Velocity Dispersion Function of Very Massive Galaxy Clusters: Abell 2029 and Coma}",
      journal = {\apjs},
         year = 2017,
        month = apr,
       volume = {229},
       number = {2},
          eid = {20},
        pages = {20},
          doi = {10.3847/1538-4365/aa653e},
archivePrefix = {arXiv},
       eprint = {1612.06428},
 primaryClass = {astro-ph.GA},
       adsurl = {https://ui.adsabs.harvard.edu/abs/2017ApJS..229...20S}
}

@ARTICLE{Sohn2022,
       author = {{Sohn}, Jubee and {Geller}, Margaret J. and {Vogelsberger}, Mark and {Damjanov}, Ivana},
        title = "{Coevolution of Brightest Cluster Galaxies and Their Host Clusters in IllustrisTNG}",
      journal = {\apj},
         year = 2022,
        month = may,
       volume = {931},
       number = {1},
          eid = {31},
        pages = {31},
          doi = {10.3847/1538-4357/ac63b7},
archivePrefix = {arXiv},
       eprint = {2201.08853},
 primaryClass = {astro-ph.CO},
       adsurl = {https://ui.adsabs.harvard.edu/abs/2022ApJ...931...31S}
}

@ARTICLE{Sohn2017b,
       author = {{Sohn}, Jubee and {Zahid}, H. Jabran and {Geller}, Margaret J.},
        title = "{The Velocity Dispersion Function for Quiescent Galaxies in the Local Universe}",
      journal = {\apj},
         year = 2017,
        month = aug,
       volume = {845},
       number = {1},
          eid = {73},
        pages = {73},
          doi = {10.3847/1538-4357/aa7de3},
archivePrefix = {arXiv},
       eprint = {1704.07843},
 primaryClass = {astro-ph.GA},
       adsurl = {https://ui.adsabs.harvard.edu/abs/2017ApJ...845...73S}
}

@ARTICLE{Sohn2024b,
       author = {{Sohn}, Jubee and {Geller}, Margaret J. and {Borrow}, Josh and {Vogelsberger}, Mark},
        title = "{The Velocity Dispersion Function for Quiescent Galaxies in Massive Clusters from IllustrisTNG}",
      journal = {\apj},
         year = 2024,
        month = oct,
       volume = {974},
       number = {1},
          eid = {26},
        pages = {26},
          doi = {10.3847/1538-4357/ad74fa},
archivePrefix = {arXiv},
       eprint = {2405.21076},
 primaryClass = {astro-ph.GA},
       adsurl = {https://ui.adsabs.harvard.edu/abs/2024ApJ...974...26S}
}

@ARTICLE{Springel2005a,
       author = {{Springel}, Volker},
        title = "{The cosmological simulation code GADGET-2}",
      journal = {\mnras},
         year = 2005,
        month = dec,
       volume = {364},
       number = {4},
        pages = {1105-1134},
          doi = {10.1111/j.1365-2966.2005.09655.x},
archivePrefix = {arXiv},
       eprint = {astro-ph/0505010},
 primaryClass = {astro-ph},
       adsurl = {https://ui.adsabs.harvard.edu/abs/2005MNRAS.364.1105S}
}

@ARTICLE{Springel2001,
       author = {{Springel}, Volker and {White}, Martin and {Hernquist}, Lars},
        title = "{Hydrodynamic Simulations of the Sunyaev-Zeldovich Effect(s)}",
      journal = {\apj},
         year = 2001,
        month = mar,
       volume = {549},
       number = {2},
        pages = {681-687},
          doi = {10.1086/319473},
archivePrefix = {arXiv},
       eprint = {astro-ph/0008133},
 primaryClass = {astro-ph},
       adsurl = {https://ui.adsabs.harvard.edu/abs/2001ApJ...549..681S}
}

@ARTICLE{Utsumi2020,
       author = {{Utsumi}, Yousuke and {Geller}, Margaret J. and {Zahid}, Harus J. and {Sohn}, Jubee and {Dell'Antonio}, Ian P. and {Kawanomoto}, Satoshi and {Komiyama}, Yutaka and {Koshida}, Shintaro and {Miyazaki}, Satoshi},
        title = "{Velocity Dispersions of Massive Quiescent Galaxies from Weak Lensing and Spectroscopy}",
      journal = {\apj},
         year = 2020,
        month = sep,
       volume = {900},
       number = {1},
          eid = {50},
        pages = {50},
          doi = {10.3847/1538-4357/aba61c},
archivePrefix = {arXiv},
       eprint = {2005.07122},
 primaryClass = {astro-ph.GA},
       adsurl = {https://ui.adsabs.harvard.edu/abs/2020ApJ...900...50U}
}

@ARTICLE{vanUitert2013,
       author = {{van Uitert}, E. and {Hoekstra}, H. and {Franx}, M. and {Gilbank}, D.~G. and {Gladders}, M.~D. and {Yee}, H.~K.~C.},
        title = "{Stellar mass versus velocity dispersion as tracers of the lensing signal around bulge-dominated galaxies}",
      journal = {\aap},
         year = 2013,
        month = jan,
       volume = {549},
          eid = {A7},
        pages = {A7},
          doi = {10.1051/0004-6361/201220439},
archivePrefix = {arXiv},
       eprint = {1211.0543},
 primaryClass = {astro-ph.CO},
       adsurl = {https://ui.adsabs.harvard.edu/abs/2013A&A...549A...7V}
}

@ARTICLE{Wagner2012,
       author = {{Wagner}, A.~Y. and {Bicknell}, G.~V. and {Umemura}, M.},
        title = "{Driving Outflows with Relativistic Jets and the Dependence of Active Galactic Nucleus Feedback Efficiency on Interstellar Medium Inhomogeneity}",
      journal = {\apj},
         year = 2012,
        month = oct,
       volume = {757},
       number = {2},
          eid = {136},
        pages = {136},
          doi = {10.1088/0004-637X/757/2/136},
archivePrefix = {arXiv},
       eprint = {1205.0542},
 primaryClass = {astro-ph.CO},
       adsurl = {https://ui.adsabs.harvard.edu/abs/2012ApJ...757..136W}
}

@ARTICLE{Wake2012,
       author = {{Wake}, David A. and {van Dokkum}, Pieter G. and {Franx}, Marijn},
        title = "{Revealing Velocity Dispersion as the Best Indicator of a Galaxy's Color, Compared to Stellar Mass, Surface Mass Density, or Morphology}",
      journal = {\apjl},
         year = 2012,
        month = jun,
       volume = {751},
       number = {2},
          eid = {L44},
        pages = {L44},
          doi = {10.1088/2041-8205/751/2/L44},
archivePrefix = {arXiv},
       eprint = {1201.4998},
 primaryClass = {astro-ph.CO},
       adsurl = {https://ui.adsabs.harvard.edu/abs/2012ApJ...751L..44W}
}

@ARTICLE{Weigel2016,
       author = {{Weigel}, Anna K. and {Schawinski}, Kevin and {Bruderer}, Claudio},
        title = "{Stellar mass functions: methods, systematics and results for the local Universe}",
      journal = {\mnras},
         year = 2016,
        month = jun,
       volume = {459},
       number = {2},
        pages = {2150-2187},
          doi = {10.1093/mnras/stw756},
archivePrefix = {arXiv},
       eprint = {1604.00008},
 primaryClass = {astro-ph.GA},
       adsurl = {https://ui.adsabs.harvard.edu/abs/2016MNRAS.459.2150W}
}

@ARTICLE{Weinberger2018,
       author = {{Weinberger}, Rainer and {Springel}, Volker and {Pakmor}, R{\"u}diger and {Nelson}, Dylan and {Genel}, Shy and {Pillepich}, Annalisa and {Vogelsberger}, Mark and {Marinacci}, Federico and {Naiman}, Jill and {Torrey}, Paul and {Hernquist}, Lars},
        title = "{Supermassive black holes and their feedback effects in the IllustrisTNG simulation}",
      journal = {\mnras},
         year = 2018,
        month = sep,
       volume = {479},
       number = {3},
        pages = {4056-4072},
          doi = {10.1093/mnras/sty1733},
archivePrefix = {arXiv},
       eprint = {1710.04659},
 primaryClass = {astro-ph.GA},
       adsurl = {https://ui.adsabs.harvard.edu/abs/2018MNRAS.479.4056W}
}

@ARTICLE{Zahid2018,
       author = {{Zahid}, H. Jabran and {Sohn}, Jubee and {Geller}, Margaret J.},
        title = "{Stellar Velocity Dispersion: Linking Quiescent Galaxies to Their Dark Matter Halos}",
      journal = {\apj},
         year = 2018,
        month = jun,
       volume = {859},
       number = {2},
          eid = {96},
        pages = {96},
          doi = {10.3847/1538-4357/aabe31},
archivePrefix = {arXiv},
       eprint = {1804.04492},
 primaryClass = {astro-ph.GA},
       adsurl = {https://ui.adsabs.harvard.edu/abs/2018ApJ...859...96Z}
}

@ARTICLE{Zahid2016,
       author = {{Zahid}, H. Jabran and {Geller}, Margaret J. and {Fabricant}, Daniel G. and {Hwang}, Ho Seong},
        title = "{The Scaling of Stellar Mass and Central Stellar Velocity Dispersion for Quiescent Galaxies at z<0.7}",
      journal = {\apj},
         year = 2016,
        month = dec,
       volume = {832},
       number = {2},
          eid = {203},
        pages = {203},
          doi = {10.3847/0004-637X/832/2/203},
archivePrefix = {arXiv},
       eprint = {1607.04275},
 primaryClass = {astro-ph.GA},
       adsurl = {https://ui.adsabs.harvard.edu/abs/2016ApJ...832..203Z}
}

@ARTICLE{Zubovas2016,
       author = {{Zubovas}, Kastytis and {King}, Andrew},
        title = "{The small observed scale of AGN-driven outflows, and inside-out disc quenching}",
      journal = {\mnras},
         year = 2016,
        month = nov,
       volume = {462},
       number = {4},
        pages = {4055-4066},
          doi = {10.1093/mnras/stw1845},
archivePrefix = {arXiv},
       eprint = {1607.07258},
 primaryClass = {astro-ph.GA},
       adsurl = {https://ui.adsabs.harvard.edu/abs/2016MNRAS.462.4055Z}
}
\end{document}